\documentclass[nobibnotes,onecolumn,letterpaper,aps,prc,
superscriptaddress,nofootinbib,showpacs,floatfix]{revtex4-1}

\usepackage{graphicx}
\usepackage{comment}
\usepackage{setspace}
\usepackage{amsmath}
\usepackage[textwidth=16cm,textheight=22cm]{geometry}
\usepackage{color}
\usepackage{indentfirst}
\usepackage{xspace}
\usepackage{verbatim}
\usepackage{epstopdf}

\usepackage{lineno}
\usepackage{dcolumn}
\usepackage{bm}
\usepackage{epsfig}
\usepackage[T1]{fontenc} 
\usepackage[pdftex,bookmarks]{hyperref}
\usepackage{color}
\usepackage[utf8]{inputenc}
\usepackage{float}
\usepackage{multirow}
\newcommand{ \be }{\begin{linenomath*}\begin{eqnarray}}
\newcommand{ \ee }{\end{eqnarray}\end{linenomath*}}
\newcommand{ \la }{\langle}

\newcommand{ \ra }{\rangle}

\def \mean#1 {{\la #1 \ra}}

\newcommand{ \pp}{pp}
\newcommand{ \pA}{p--A}
\newcommand{ \pPb}{p--Pb}
\newcommand{ \PbPb}{Pb--Pb}
\newcommand{ \AonA}{A--A}

\newcommand{ \pt }{\textit{p}_{\rm T}}
\newcommand{ \gevc }{\rm{GeV}/\textit{c}}

\newcommand{ \ptOne}{\textit{p}_{\rm T,1}}
\newcommand{ \ptTwo}{\textit{p}_{\rm T,2}}
\newcommand{ \pti}{\textit{p}_{\rm T,i}}

\newcommand{ \dpt }{\rm{d}\textit{p}_{\rm T}}

\newcommand{ \DeltaPtPt}{\langle  \Delta \pt \Delta \pt \rangle}

\newcommand{ \dphi}{\rm{d}\varphi}
\newcommand{ \phiOne}{\varphi_1}
\newcommand{ \phiTwo}{\varphi_2}

\newcommand{ \dphiOne}{\rm{d}\varphi_1}
\newcommand{ \dphiTwo}{\rm{d}\varphi_2}
\newcommand{ \Dphi }{\Delta \varphi}

\newcommand{ \etaOne }{\eta_1}
\newcommand{ \etaTwo }{\eta_2}

\newcommand{ \deta }{\rm{d}\eta}
\newcommand{ \detaOne }{\rm{d}\eta_1}
\newcommand{ \detaTwo }{\rm{d}\eta_2}
\newcommand{ \Deta }{\Delta \eta}
\newcommand{ \etaPhi }{\left( \eta,\varphi \right)}

\newcommand{ \DetaDphi }{\left(  \Delta \eta, \Delta \varphi \right)}
\newcommand{ \rhoOne}{\rho_1}
\newcommand{ \rhoTwo}{\rho_2}
\newcommand{\Rtwo}{\rm{R_2}}
\newcommand{\Ptwo}{\rm{P_2}}
\newcommand{\RtwoLS}{\rm{R_2^{LS}}}
\newcommand{\RtwoUS}{\rm{R_2^{US}}}
\newcommand{\RtwoCI}{\rm{R_2^{CI}}}
\newcommand{\RtwoCD}{\rm{R_2^{CD}}}

\newcommand{\PtwoCI}{\rm{P_2^{CI}}}
\newcommand{\PtwoCD}{\rm{P_2^{CD}}}

\definecolor{dgreen}{cmyk}{1.,0.,1.,0.2}        
\definecolor{orange}{cmyk}{0.,0.353,1.,0.}    

\def \s {\sqrt{\textit{s}}}

\long\def\/*#1*/{}

\begin{document}
\date{\today}  
\title{Simulation studies of $\Rtwo \DetaDphi$ and $\Ptwo \DetaDphi$
correlation functions in \pp\ collisions with the PYTHIA and HERWIG models}
\author{Baidyanath Sahoo}
\email{baidya@iitb.ac.in}
\author{Basanta Kumar Nandi}
\affiliation{Department of Physics, Indian Institute of Technology
 Bombay, Mumbai - 400076, India }
\author{Prabhat Pujahari}
\affiliation{Department of Physics, Indian Institute of Technology
 Madras, Chennai - 600025, India }
\author{Sumit Basu}
\email{sumit.basu@cern.ch}
\affiliation{Department of Physics and Astronomy, Wayne State
 University, Detroit, 48201 USA}
\author{Claude Pruneau}
\email{claude.pruneau@wayne.edu}
\affiliation{Department of Physics and Astronomy, Wayne State
 University, Detroit, 48201 USA}

\begin{abstract}
We report studies of charge-independent (CI) and charge-dependent (CD) two-particle differential-number
correlation functions, $\Rtwo\DetaDphi$, and transverse momentum ($\pt$) correlation functions,
$\Ptwo \DetaDphi$, of charged particles in $\s$ = 2.76 TeV pp collisions with the PYTHIA and HERWIG models.
Model predictions  are presented for inclusive charged hadrons ($h^\pm$), as well as pions ($\pi^\pm$), kaons
(K$^\pm$), and (anti-)protons ($\rm \bar{p}$/p) in the ranges $0.2 < \pt \le 2.0~\gevc$, $2.0 < \pt \le 5.0~\gevc$,
and $5.0 < \pt \le 30.0~\gevc$, with full azimuthal coverage in the 
range $|\eta|< 1.0$. We compare the model predictions for the strength
and shape of the $\Rtwo$ and $\Ptwo$ correlators as these
pertain to recent measurements by the ALICE
collaboration. The $\Rtwo$ and $\Ptwo$ correlation functions
estimated with PYTHIA and HERWIG exhibit qualitatively similar
near-side and away-side correlation structures
but feature important differences. Our analysis
indicates that comparative studies of $\Rtwo$ and $\Ptwo$
correlation functions would provide valuable insight towards
the understanding of particle production in pp collisions, and by
extension, should also be useful in studies of heavy-ion collisions. Comparison of the $\Deta$ dependence of $\Rtwo$ and $\Ptwo$
could contribute, in particular, to a better
understanding and modeling of the angular ordering of particles
produced by hadronization in jets, as well as a better description of
jet fragmentation functions of identified species at low momentum fraction $(z)$.

\end{abstract}


\maketitle

\section{Introduction}
\label{sec:introduction}

Measurements of integral and differential correlation functions
constitute essential tools for the study of proton-proton (\pp) and
heavy-ion (\AonA) collisions at relativistic energies. Two- and
multi-particle azimuthal correlations functions have provided
evidence for the existence of anisotropic flow in \AonA\
collisions~\cite{Adams:2005dq,Adcox:2004mh,Aamodt:2010pa,Aamodt:2011by, CMS:2013bza,CMS-harmonic,ATLAS-harmonic}, quark scaling (approximate) of flow coefficients in \AonA\ collisions at RHIC and
LHC~\cite{PhysRevLett.98.162301,Abelev:2014pua,Adam:2016nfo,ATLAS:2012at}. They were also used to investigate the presence of flow in smaller
systems (e.g.,  \pA\ and high multiplicity \pp\ collisions)~\cite{Esumi:2017xvf,
 MalgorzataALICE:2017Corr,X.Zhu:2013JetCorrelation,Abelev:2014mda,
 Bernardes:2017eup,Sirunyan:2018toe}. Differential two-particle
(number) correlation functions additionally enabled the discovery of jet quenching at RHIC~\cite{Adler:2002tq,PhysRevC.77.011901,PhysRevC.77.011901}
and its detailed characterization in \AonA\ collisions at both RHIC
and LHC~\cite{Chatrchyan:2011sx}. Several other correlation
functions, including number and transverse momentum correlation
functions~\cite{Adams:2005ka,Agakishiev:2011fs} have been measured and 
investigated both at RHIC and LHC to better understand the particle
production dynamics and study the properties of the matter produced in
\pp\ and \AonA\ collisions~\cite{S.PrattPRL:2000BalFun1st,
 Pratt:2015jsa,Aggarwal:2010ya,Adams:2003kg,Abelev:2013csa,Abelev:2013csa,
 AliceDptDptLongPaper}. Among these, the recent measurements of number
correlation, $\Rtwo$, and differential transverse momentum
correlation, $\Ptwo$, defined in Sec.~\ref{sec:definition}, have
enabled independent confirmation of the collective nature of the
azimuthal correlations observed in \PbPb\
collisions~\cite{Adam:2017ucq}, as well as the identification of
noticeable differences in the $\Deta$ and $\Dphi$ dependence of these
correlation functions~\cite{AliceDptDptLongPaper}. These measurements
show that the near-side peak of both CI and CD correlations is
significantly narrower, at any given \AonA\ collision centrality in
$\Ptwo$ than in $\Rtwo$ correlation functions. This confirms
~\cite{Sharma:2008qr} that comparative measurements of
$\Ptwo$ and $\Rtwo$ correlation functions may provide additional
sensitivity to the underlying particle production mechanisms in heavy-ion collisions.
In this work, we seek to establish whether the difference observed in
\cite{AliceDptDptLongPaper} can be readily explained by jet
contributions. To this end, we examine predictions of the $\Rtwo$
and $\Ptwo$ correlation functions by the
PYTHIA~\cite{P.SkandsJHEP:2006} and HERWIG 
\cite{G.CorcellaJHEP:2001Herwig6.5Manual} models that are known to
quantitatively reproduce many jet related observables reported and
compiled by RHIC and LHC experiments
~\cite{jetQCDBySeymour,jetPhyRhicLhc,jetCrossSecInpp}. We examine the differential
correlation functions $\Rtwo$ and $\Ptwo$ in \pp\ collisions with a particular focus on
particles produced in the range $0.2 < \pt \le 2.0$ \gevc\ reported by
ALICE~\cite{Adam:2017ucq,AliceDptDptLongPaper} but also extend our
study to include higher momentum ranges to further examine how
the two observables behave for higher particle momenta expected to be
dominated by jet production.

Particle production in high-energy nucleus-nucleus collisions is
governed by several conservation laws including (electric) charge
conservation, baryon number conservation, strangeness conservation,
as well as energy-momentum conservation. At very large collision
energy, the yield of anti-particles and particles are nearly equal.
Limited information is thus gained by studying the yields, e.g., $\pi^+$
and $\pi^-$ individually. Additional insight may be provided,
however, by comparative studies of like-sign (LS) and  unlike-sign (US)
particle pairs, e.g., $\pi^+,\pi^+$ and $\pi^+,\pi^-$, or
baryon-baryon and baryon-anti-baryon particle pairs. We thus study
predictions of the models for both charge-independent (CI) and
charge-dependent (CD) pair combinations.

This paper is organized as follows. Section~\ref{sec:definition}
presents definitions of the $\Rtwo$ and $\Ptwo$ correlation
functions studied in this work and describes how they are computed. 
The PYTHIA and HERWIG models, and the conditions under which they
were used to generate \pp\ events, are briefly described in
Sec.~\ref{sec:models}. Predictions by the models for $ \Rtwo$ and
$\Ptwo$ correlation functions are presented in Sec.
\ref{sec:Results} and conclusions are summarized in Sec. \ref{conclusion}.

\section{Correlation functions definition} 
\label{sec:definition}

The $\Rtwo$ and $\Ptwo$ correlation functions are defined in terms of
single- and two-particle densities expressed as functions
of the particle pseudo-rapidity $\eta$ and azimuthal angle 
$\varphi$
\be
\rhoOne \etaPhi &=&\frac{1}{\sigma_1} \frac{\rm{d}^2\sigma_1}{\deta \dphi}, \\ 
\rhoTwo(\etaOne, \phiOne,\etaTwo, \phiTwo)&
=& \frac{1}{\sigma_2} 
\frac{\rm{d}^4\sigma_2}{\detaOne\dphiOne\detaTwo \dphiTwo}, 
\ee
where $\sigma_1$ and $\sigma_2$ are single and two-particle
cross-sections, respectively. The correlator $\Rtwo$ is defined 
as a two-particle cumulant normalized by the product of single-particle densities
(hereafter called  normalized two-particle cumulant) according to
\be
\Rtwo (\etaOne, \phiOne,\etaTwo, \phiTwo)&
=& \frac{\rhoTwo(\etaOne, \phiOne,\etaTwo, \phiTwo) }{ \rhoOne 
 (\etaOne, \phiOne)\rhoOne (\etaTwo, \phiTwo)}-1,
\ee
while the $\Ptwo$ correlation function is defined in
terms of the differential correlator $\DeltaPtPt$
normalized by the square of the average
transverse momentum, $\pt$, to make it dimensionless, as follows
\be
\Ptwo (\etaOne, \phiOne,\etaTwo, \phiTwo)&
=&\frac{\DeltaPtPt(\etaOne, \phiOne,\eta_2, \phiTwo)}
{\la \pt\ra^2}.
\ee
The $\la \Delta \pt\Delta \pt \ra$ differential correlator~\cite{Sharma:2008qr} is defined
according to 
\begin{equation}
\la \Delta \pt\Delta \pt \ra(\etaOne, \phiOne,\eta_2, \phiTwo) = 
\frac{  \int_{p_{\rm T,\min}}^{p_{\rm T,\max}} \Delta p_{\rm T,1} 
 \Delta p_{\rm T,2}  \  \rho_2({\vec p}_{1},{\vec p}_{2}) \ {\rm d}p_{\rm T,1}{\rm d}p_{\rm T,2}  
  } 
{\int_{p_{\rm T,\min}}^{p_{\rm T,\max}} \rho_2({\vec p}_{1},{\vec p}_{2}) \ {\rm d}p_{\rm T,1}{\rm d}p_{\rm T,2} }
\end{equation}
where $\Delta\pti = \pti - \la \pt \ra$ and $\la \pt \ra$ is the
inclusive mean transverse momentum  
\be
\la \pt \ra = \frac{\int_{p_{\rm T,\min}}^{p_{\rm T,\max}}\rhoOne \pt \dpt} 
{\int_{p_{\rm T,\min}}^{p_{\rm T,\max}}\rhoOne \dpt}.
\ee
In addition to its sensitivity to the presence of particle
correlations, $\Ptwo$ is also determined by the momentum of the
correlated particles~\cite{Adam:2017ucq}. It is positive whenever particle pairs
emitted at specific azimuthal angle and pseudo-rapidity
differences are more likely to both have transverse momenta higher (or
lower) than the $\langle \pt \rangle$ and negative when a high-$\pt $
particle ($ \pt > \langle \pt \rangle$) is more likely to be
accompanied by a low-$\pt $ particle ($ \pt < \langle \pt
\rangle$). For instance, particles emitted within a jet typically have
higher $\pt $ than the inclusive average. Jets, therefore, contribute a large positive value to $\Ptwo$. Hanbury-Brown and Twiss (HBT) correlations, determined by
pairs of identical particles with $\ptOne \approx \ptTwo$ should
likewise contribute positively to this correlator. However, particle
production involving a mix of low- and high-momenta correlated
particles can contribute both positively and negatively. Based on this simple
observation, one expects the internal structure of jets to influence the $\Delta\eta$, $\Delta\varphi$ dependence of the near-side peak of $\Ptwo$ correlation functions.
\begin{figure}[ht!]
\begin{center} 
\includegraphics[scale=0.25]{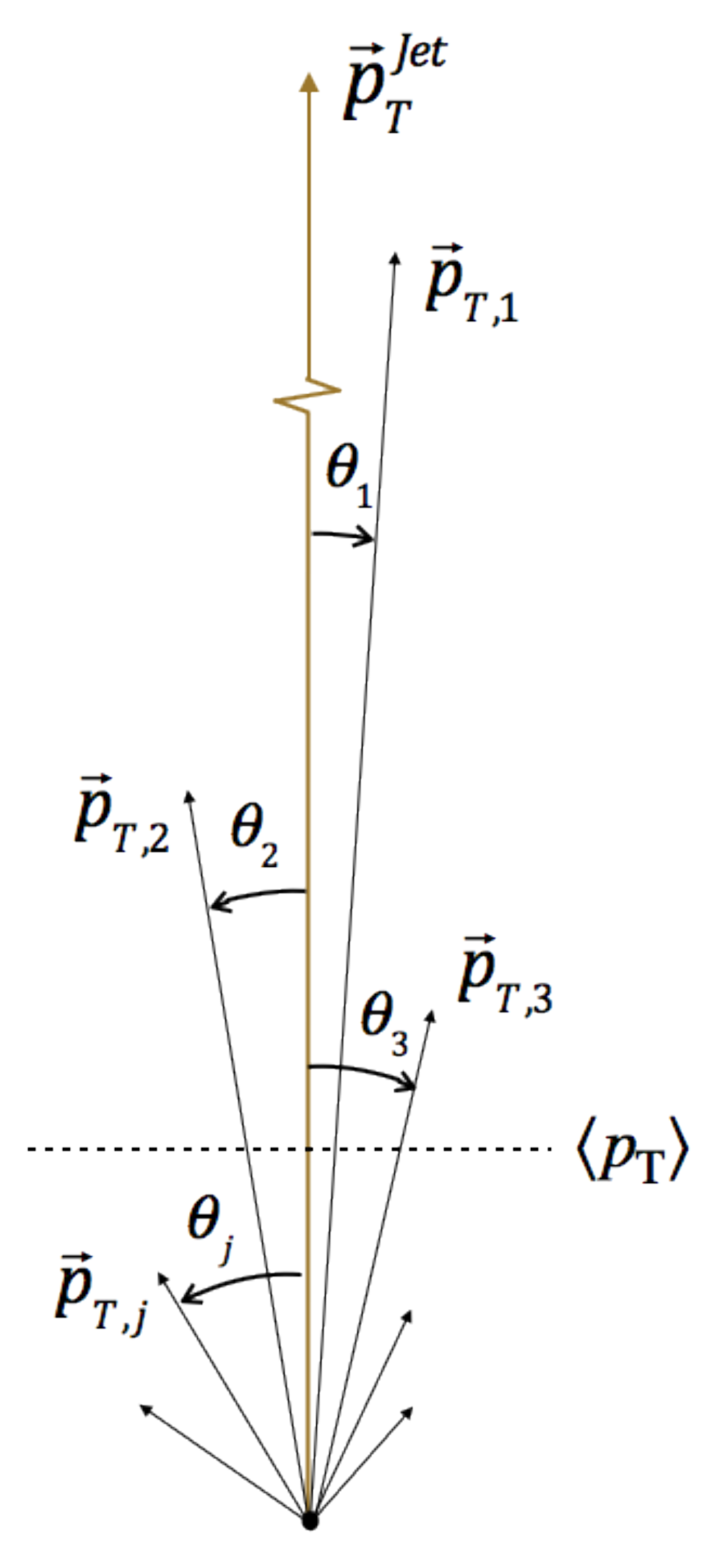}
\caption{Schematic representation of the transverse momentum and angular ordering of jet constituents relative to the jet axis, ${\vec p}_{\rm T}^{\hspace{0.01in} \it Jet}$.}
\label{fig:jetFigure}
\end{center} 
\end{figure} 
Within jets, as schematically illustrated in Fig.~\ref{fig:jetFigure}, high-$\pt$ particles are
predominantly emitted at small polar angles relative to the jet axis, while lower $\pt$
particles span a larger angular range. This leads to an effective ordering of the particles
(typical) $\pt$ relative to the polar angle $\theta$. In turn, this also leads to an effective
$\pt$ ordering in the $\Delta\eta$ vs. $\Delta\varphi$ plane. For instance, one expects
that for small $\Delta\eta$, $\Delta\varphi$ separations, high-$\pt$ particles (i.e., $p_{{\rm T}, i}\gg \la \pt\ra$) should dominate the $\Ptwo$
correlation and contribute positive $\Delta \pt\Delta \pt$ values. Likewise, at very large
$\Delta\eta$, $\Delta\varphi$ separation, the correlation strength should be determined
by particle pairs with $p_{{\rm T}, i} < \la \pt\ra$ thereby also yielding positive
$\Delta \pt\Delta \pt$ values. However, there shall also be an intermediate
$\Delta\eta$, $\Delta\varphi$ range such that
pairs consist of one high-$\pt$ particle and one low-$\pt$ particle yielding negative
$\Delta \pt\Delta \pt$ values on average. This should thus produce a narrowing and
possibly a non-monotonic $\Delta\eta$, $\Delta\varphi$ dependence of the correlation
strength. We shall demonstrate, in the following, that both PYTHIA and HERWIG do in
fact exhibit such behavior.

In this work, the correlators $\Rtwo$ and $\Ptwo$ are reported
as function of the differences $\Deta= \etaOne - \etaTwo$ and
$\Dphi=\phiOne - \phiTwo$ by averaging across the mean pseudo-rapidity
$\bar \eta = \frac{1}{2}(\etaOne + \etaTwo)$ and the mean azimuthal angle
$\bar \varphi = \frac{1}{2}(\phiOne + \phiTwo)$ acceptance
according to
\be
\rm{O}(\Deta, \Dphi) 
= \frac{1}{\Omega(\Deta)} \int \rm{O}(\etaOne, \phiOne,\etaTwo, 
\phiTwo) \delta(\Dphi - \phiOne + \phiTwo) \dphiOne \dphiTwo \\ 
\nonumber 
\times \delta(\Deta - \etaOne + \etaTwo) \detaOne \detaTwo,
\ee
where $\Omega(\Deta)$ represents the width of the acceptance in
$\bar{\eta}$ at a given value of $\Deta$ and angle differences
$\Dphi$ are calculated modulo $2\pi$ and shifted by $-\pi/2$ for
convenience of representation in the figures. The analysis of the
$\Rtwo$ and $\Ptwo$ correlation functions are carried out for charge
combination pairs $(+-)$, $(-+)$, $(++)$, and $(--)$
separately. Like-sign pairs correlations are averaged to yield LS
correlations, $\rm{O}^{\rm LS} = \frac{1}{2}[ \rm{O}^{++} + \rm{O}^{--}]$, and US
correlators are obtained by averaging $(+-)$ and $(-+)$ correlations,
$\rm{O}^{\rm US} = \frac{1}{2}[ \rm{O}^{+-} + \rm{O}^{-+}]$. The LS and US correlations are then combined to yield charge-independent and charge-dependent
correlation functions according to
\be
\label{eq:CI}
\rm{O}^{\rm CI} &=& \frac{1}{2}\left[\rm{O}^{\rm US} + \rm{O}^{\rm LS}\right],\\
\label{eq:CD}
\rm{O}^{\rm CD} &=& \frac{1}{2} \left[\rm{O}^{\rm US} - \rm{O}^{\rm LS}\right].
\ee 
The CI correlation function measures the
average of all correlations between charged particles while the CD
correlation function is sensitive to the difference of US and
LS pairs and is largely driven, as such, by charge conservation
effects. The CD correlation function is proportional to the
charge balance function~\cite{S.PrattPRL:2000BalFun1st} when the
yields of positive and negative particles are equal~\cite{C.PruneauPRC:2002Fluct}.\\

We repeated the analysis and used the sub-sampling technique to obtain a more accurate estimation of statistical uncertainty. The Monte Carlo data sample was divided into 10 segments of equal size— e.g. equal number of events. Each sub-sample was analyzed independently. We then extracted the mean values and calculated the sample standard deviations ($\sigma$) according to
\be
\label{eq:stdDev}
\sigma = \sqrt{\frac{\sum_{i}(O_{i} -<O>)^2}{N-1}}
\ee 
where “N-1” used instead of “N” depending on Bessel’s correction and , where i= 1,2,…10. 
  The error on the mean is calculated bin-by-bin using the general formula $\sigma_{error} = \frac{\sigma}{\sqrt{N}}$ .

\section{Monte Carlo Models} 
\label{sec:models}

The impact of jet production on $\Rtwo$ and $\Ptwo$ correlation
functions in \pp\ is studied with Monte Carlo simulations carried
out with the event generators PYTHIA 6.425, tune Perugia-0
~\cite{P.SkandsJHEP:2006,P.SkandsPRD:2010PythiaTune,
 P.SkandsEPJC:2005PhysicsModel,P.Skands:2006LEP,A.BuckleyEPJC:2010LEP},
and HERWIG 6.5~\cite{G.CorcellaJHEP:2001Herwig6.5Manual}. PYTHIA and
HERWIG are both based on QCD at Leading Order (LO) but use different
parton production and hadronization schemes. PYTHIA uses the Lund
string fragmentation model for high-$\pt$ parton
hadronization while the production of soft particles (i.e., the
underlying event) is handled through fragmentation of mini-jets from initial and
final state radiation, multiple parton interactions (MPI), and proton
remnants~\cite{B.AnderssonPR:1983FragmentString}. The kPyJets process
responsible for jet production uses the CTEQ6l~\cite{J.PumplinJHEP:2000CTEQ6L}
parametrization of the proton parton distribution function (PDF) tuned
for LHC energies. HERWIG events were generated based on the jet
generation process 1500~\cite{G.CorcellaJHEP:2001Herwig6.5Manual} and
the CTEQ5L~\cite{H.L.LaiEPJC:2000CTEQ5L} parametrization of the proton
PDFs with hard color-singlet exchange between two partons
~\cite{A.H.MuellerPRB:1992HerwigFragmentCluster} using
leading-logarithmic (LL) BFKL~\cite{GavinAPPB:1999BFKL} calculations
in ALICE environment. In HERWIG, the perturbative parton evolution
ends with the production of clusters subsequently decayed into final-state hadrons.

In order to study the correlation functions with reliable statistical accuracy,
and given the jet production cross-section falls steeply with
increasing transverse momentum, we generated equal number of PYTHIA and
HERWIG events in three hard QCD ($2\rightarrow 2$ processes) $\hat \pt$ bins:
5.0 - 10.0 \gevc, 10.0 - 20.0\gevc, and 20.0 - 30.0 \gevc
~\cite{LHC-Diaz:HighPtPythiaHerwig}. A total of $2\times 10^8$
events were generated with PYTHIA and $2\times 10^8$ HERWIG events
were produced in each ${\hat p}_{\rm T}$ bins. Single- and
two-particle densities were calculated independently in each ${\hat p}_{\rm T}$ bin and
averaged with weights corresponding to their respective fractional cross-sections.

Charge, baryon number, or strangeness balance function~\cite{S.PrattPRL:2000BalFun1st} are of interest to study the role of
conservation laws and the dynamics of particle transport in elementary
and heavy-ion collisions. However, charge balance functions should be
proportional and thus equivalent to the correlator $\RtwoCD$
provided the measured multiplicities of positively and negatively
charged particles are equal. We verified the applicability of the
equivalence by comparing the differential cross-sections of positively
and negatively charged hadrons, $h^{\pm}$,
and found that the ratio of cross-sections is
of order unity in the $\pt$ range of interest of this study. It is
then legitimate to use the $\RtwoCD$ correlation function as
a proxy for the balance function: balance functions are thus not
explicitly reported in this study.

\section{Model predictions} 
\label{sec:Results}

We begin with a discussion of unidentified charged hadron correlation
functions in sec.~\ref{sec:ResultsInclusive}. Correlation functions
for identified particles, e.g., pions, kaons, and protons, are
presented in sec.~\ref{sec:ResultsPID}.

\subsection{Inclusive charged hadron correlations} 
\label{sec:ResultsInclusive}

\begin{figure}[htb] 
 \begin{center} 
  \includegraphics[scale=0.81]{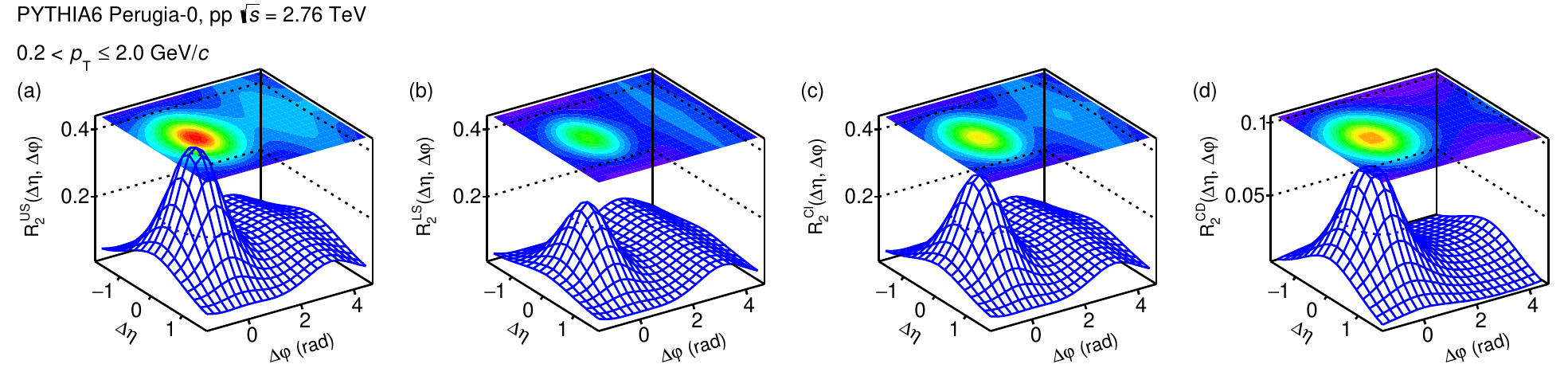}
  \caption{Normalized two-particle cumulants (a) $\RtwoUS$,
   (b) $\RtwoLS$, (c) $\RtwoCI$, (d) $\RtwoCD$
   obtained from PYTHIA simulations of \pp\
   collisions at $\s$ = 2.76 TeV for charged hadrons in the
   pseudo-rapidity range $|\eta| < 1.0$ and the transverse
   momentum range $0.2 < \pt \leq 2.0$ \gevc.}
  \label{fig:cR2UsLs}
 \end{center} 
\end{figure}

We focus our discussion on CI and CD correlation
functions. Associated balance functions, $\rm B$, can be obtained by 
multiplying $\RtwoCD$ correlation functions with integrals of
the hadron cross-sections.
Figure~\ref{fig:cR2UsLs} illustrates the calculation of CI and CD
correlation functions based on the correlators $\RtwoUS$ and
$\RtwoLS$. Panels (a,b) present examples of these correlation
functions calculated with PYTHIA for particles in the transverse momentum range
$0.2 < \pt \le 2.0$ \gevc\ and pseudo-rapidity range $|\eta|<
1.0$. The US and LS correlation functions are combined according to
Eqs.~(\ref{eq:CI}) and (\ref{eq:CD}) to obtain CI and CD correlation
functions shown in panels (c,d) of the same figure.

The US and LS
correlation functions both feature a prominent near-side peak centered
at $(\Deta,\Dphi)=(0,0)$. One notes, however, that the peak observed
in US correlation functions is taller than that observed in LS
correlation functions. This leads to a modest and narrow near-side
peak in the CD correlation function shown in panel (d). The amplitude
and shape of this peak are determined by the (charge) pair production and
hadronization processes. We show in this article that PYTHIA and HERWIG make
quantitatively different predictions of these features. Measurements of $\RtwoCD$
correlation functions shall thus provide a valuable basis to test the underlying
mechanisms used in these models for $\rm q\bar{\rm q}$ pair creation
and hadronization of partons into hadrons, $q(\bar{q}) \rightarrow h^{\pm}$.

\subsubsection{Charge Independent Correlations} 
\label{sec:ChargeIndependentCorrelations}
\begin{figure}[htb] 
 \begin{center} 
  \includegraphics[scale=0.8]{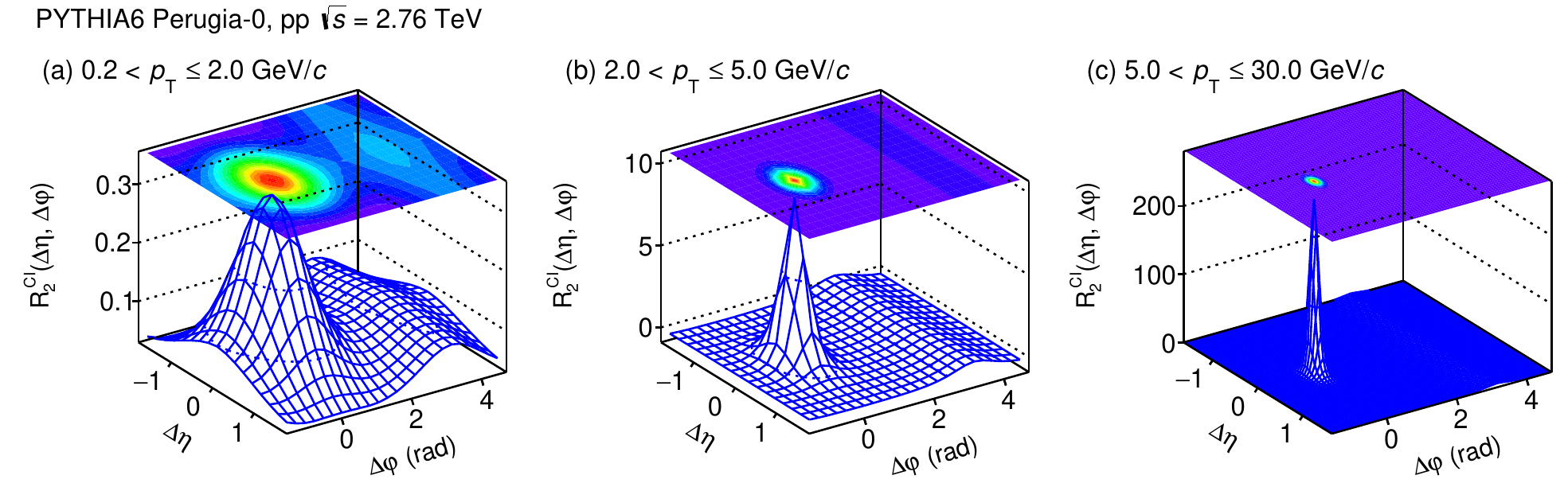}
  \includegraphics[scale=0.8]{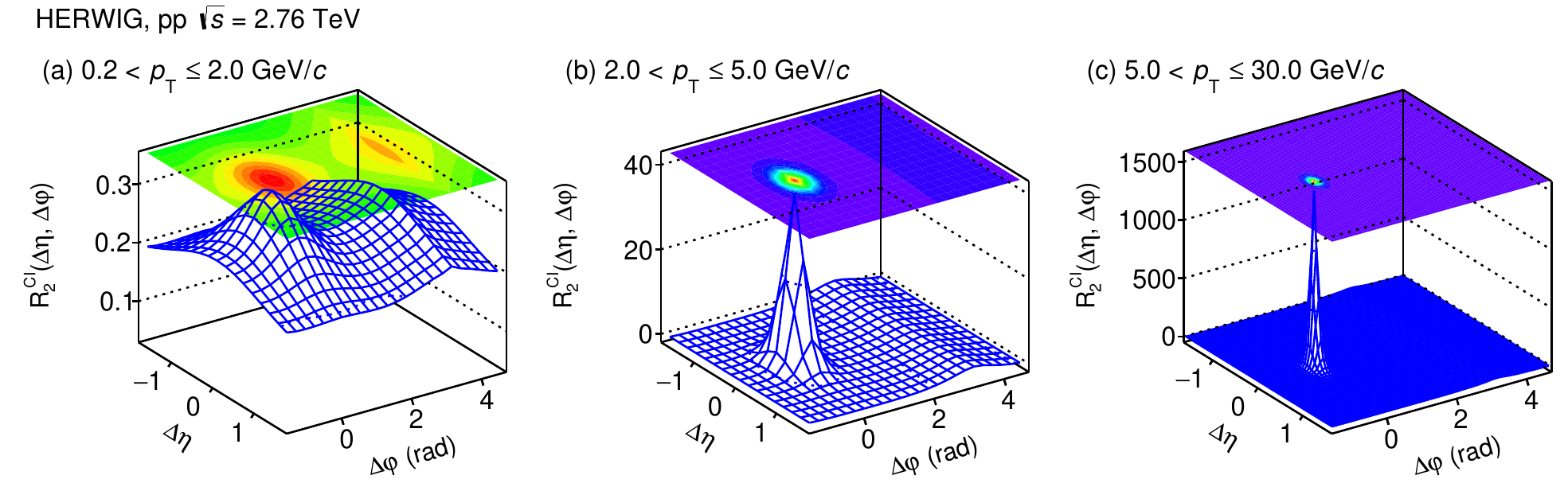}
  \caption{Correlation functions $\RtwoCI$ of charged
   hadrons, in selected $\pt $ ranges, obtained with PYTHIA 
   (top panel) and HERWIG (bottom panel) in \pp\ collisions at $\s$ = 2.76 TeV.}
  \label{fig:cR2CI}
 \end{center} 
\end{figure}

\begin{figure}[htb] 
 \includegraphics[scale=0.8]{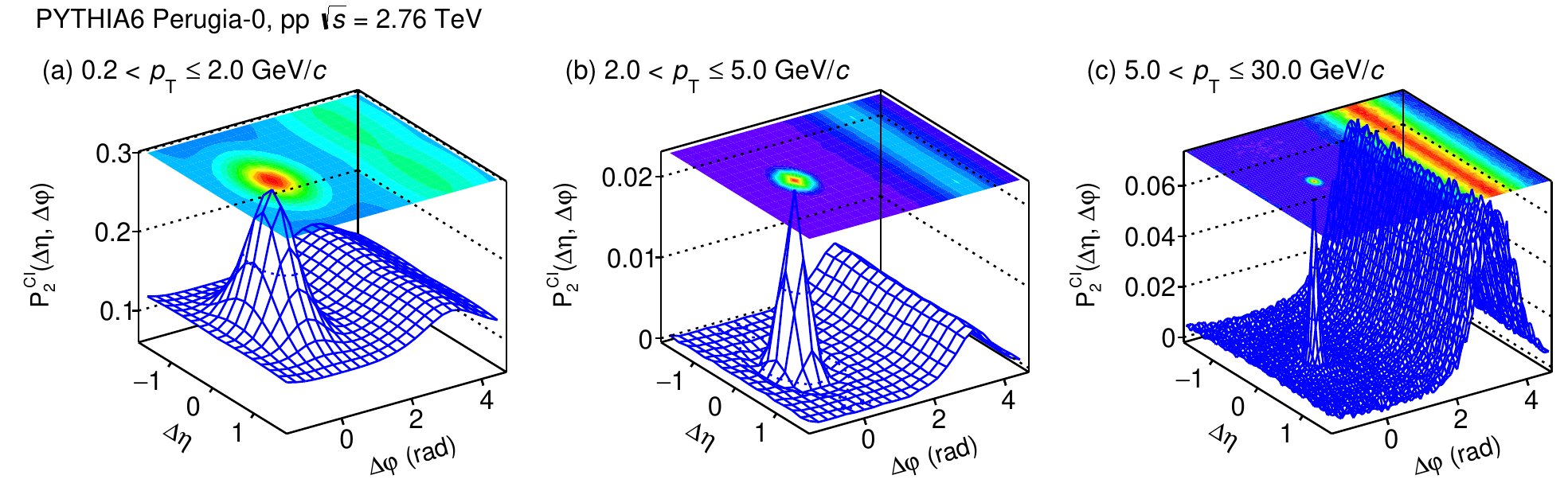}
 \includegraphics[scale=0.8]{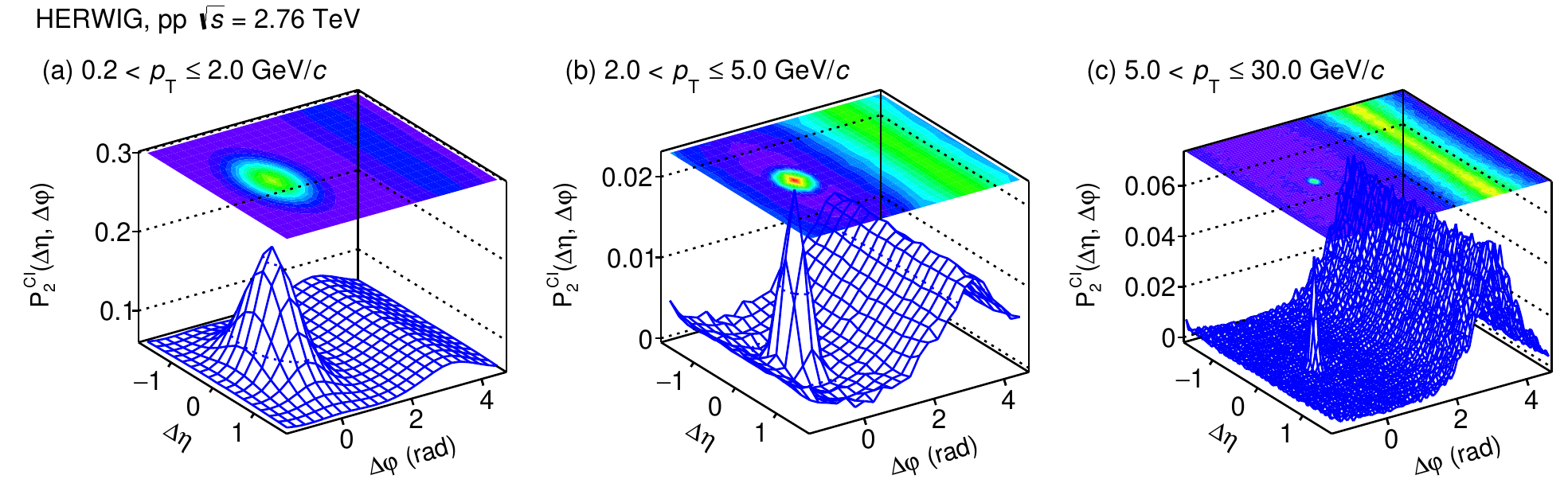}
 \caption{Correlation functions $\PtwoCI$ of charged hadrons,
  in selected $\pt$ ranges, obtained with PYTHIA (top panel) and
  HERWIG (bottom panel) in \pp\ collisions at $\s$ = 2.76 TeV.}
 \label{fig:cP2CI}
\end{figure}

We first compare the $\Rtwo$ and $\Ptwo$ correlation functions for
CI charge combinations obtained in simulations of \pp\ collisions
with the PYTHIA and HERWIG generators. Figures~\ref{fig:cR2CI} and
\ref{fig:cP2CI} present the $\RtwoCI$ and $\PtwoCI$ correlation functions,
respectively.


\begin{figure}[ht!]
 \hspace*{-1.3cm}%
 \includegraphics[scale=0.4]{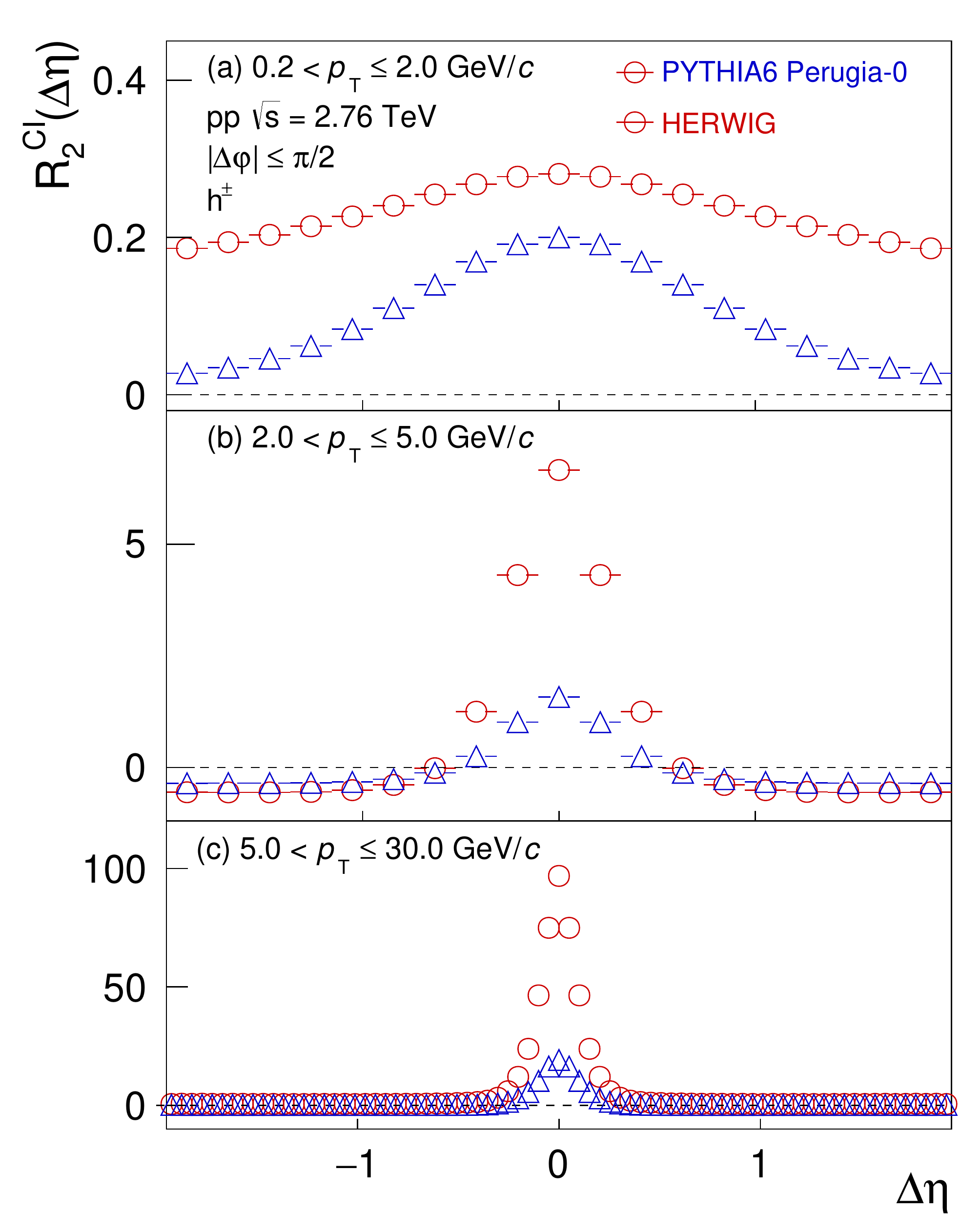}
\includegraphics[scale=0.4]{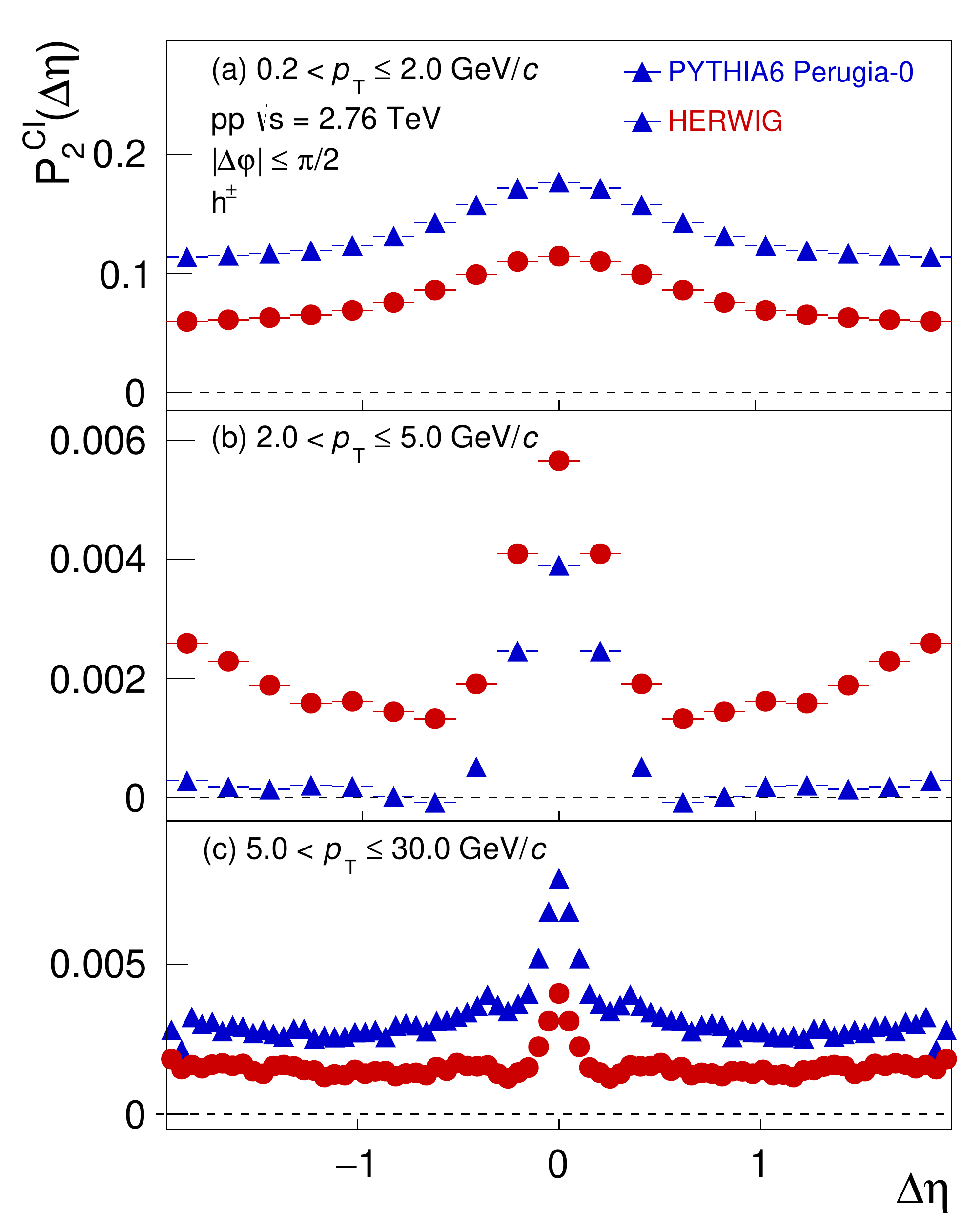}

 \caption{Projections onto $\Deta$ of the $\RtwoCI$ (left
  panel) and $\PtwoCI$ (right panel) correlation functions
  calculated with PYTHIA (blue) and HERWIG (red) for $h^{\pm}$ in
  \pp\ collisions at $\s$ = 2.76 TeV in selected $\pt$ ranges. The
  $\Deta$ projections are calculated as averages of the
  two-dimensional correlations in the ranges $|\Dphi| \leq \pi/2$.}
 \label{fig:cDetaCI}
\end{figure} 

\begin{figure}[ht!]
 \hspace*{-1.3cm}%
 \includegraphics[scale=0.4]{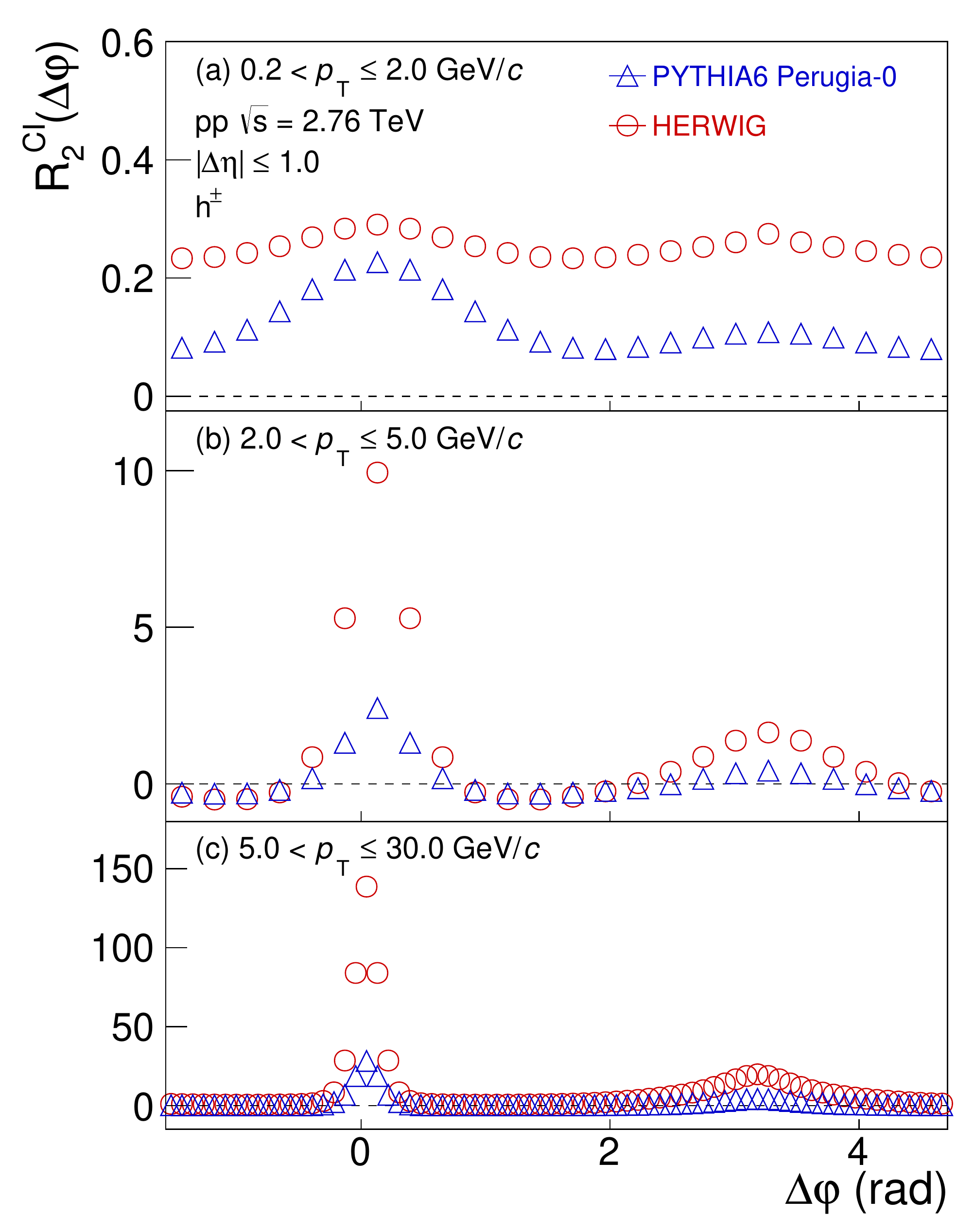}
 \includegraphics[scale=0.4]{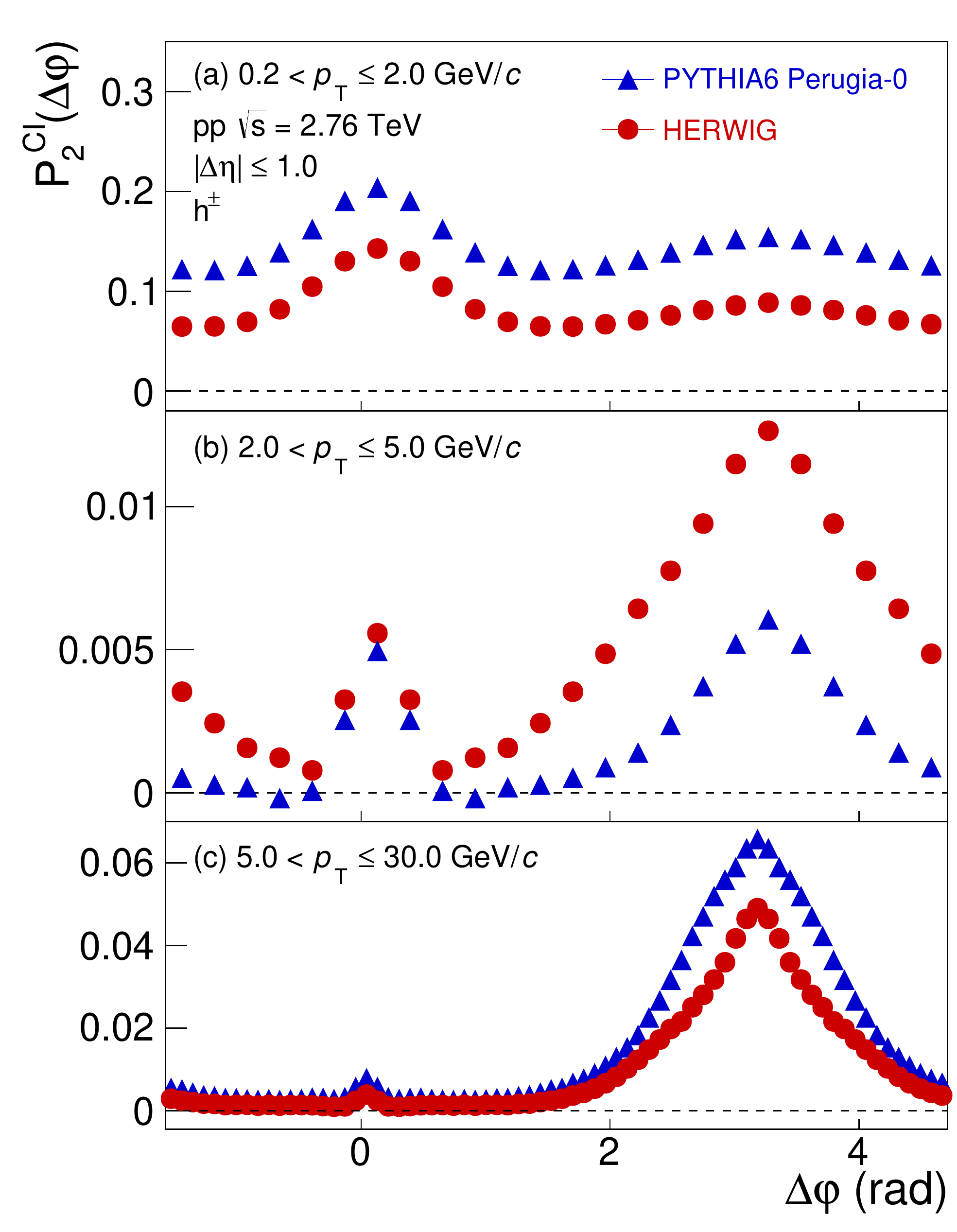}
 \caption{Projections onto $\Dphi$ of $\RtwoCI$ (left panel)
  and $\PtwoCI$ (right panel) correlation functions
  calculated with PYTHIA (blue) and HERWIG (red) for $h^{\pm}$ in
  \pp\ collisions at $\s$ = 2.76 TeV in selected $\pt$ ranges. The
  $\Dphi$ projections are calculated as averages of the
  two-dimensional correlations in the ranges $|\Deta| \leq 1.0$.}
 \label{fig:cDphiCI}
\end{figure} 

Correlation functions are presented for unidentified
charged hadrons calculated in momentum ranges: (i) $0.2 < \pt
\le 2.0$ \gevc, (ii) $2.0 < \pt \le 5.0$ \gevc, and
(iii) $5.0 < \pt \le 30.0$ \gevc. In these and the following figures, the $\la \pt\ra$ values
used for the calculation of $\Delta \pt\Delta \pt$ are the $\pt$ averages of particles
produced in each of these three ranges, respectively.
The first momentum range samples the underlying
event in \pp\ collisions and is relevant for comparisons with bulk
particle production in \AonA\ collisions. The second range corresponds to the
coalescence range~\cite{0954-3899-25-2-020,Molnar:2003ff}, while the
third range shifts the focus on particles produced by jet
fragmentation. We find calculations of the $\Rtwo$ and $\Ptwo$ correlation
functions in these three ranges yield qualitatively similar results. However, they
also exhibit interesting quantitative differences which we discuss in details, in the
following, based on projections onto the $\Deta$ and $\Dphi$ axes.
The $\RtwoCI$ and $\PtwoCI$ correlation functions feature a prominent near-side
peak centered at $(\Deta,\Dphi)=(0,0)$ as well as an away-side structure, centered at
$\Dphi=\pi$, and extending across the range $|\Deta|\le 1.6$ (truncated in the
figure to avoid the fluctuations at larger $\Deta$). Such near-side and away-side
features have been observed in triggered and non-triggered
correlation function measured in a variety of collision systems and beam energies
~\cite{Adam:2017ucq,AliceDptDptLongPaper,PhysRevC.77.011901,
 X.Zhu:2013JetCorrelation,PhysRevC.75.034901,Abelev:2009af,Adare:2008ae}. In
this work, we study the predictions of the PYTHIA and HERWIG models
relative to their dependence on the particle momenta, the particle
species, and we focus, in particular, on the differences between
$\Rtwo$ and $\Ptwo$ correlation functions.

In Figs.~\ref{fig:cR2CI} and \ref{fig:cP2CI}, one observes the
longitudinal and azimuthal widths of the near-side peak of the
$\RtwoCI$ and $\PtwoCI$ correlation functions predicted
by the two models decrease monotonically with the $\pt$ range
of the particles. Additionally, the near-side peak of $\PtwoCI$
 correlations are systematically narrower than those observed in
the $\RtwoCI$ correlation functions, as reported by the ALICE
collaboration~\cite{Adam:2017ucq,AliceDptDptLongPaper}. These
differences are studied quantitatively based on the projections of the
correlation functions onto the $\Deta$ and $\Dphi$ axes presented in
Figs.~\ref{fig:cDetaCI} and \ref{fig:cDphiCI}.

Projections of the $\RtwoCI$ and $\PtwoCI$ correlation
functions onto the $\Deta$ axis calculated with the PYTHIA and HERWIG
models are presented in the left and right panels of Fig.~\ref{fig:cDetaCI},
respectively, for the three $\pt$ ranges already considered. One
observes that the models make quantitatively different predictions
for both correlation functions in all three $\pt$
ranges. Indeed, both models yield peaks centered at ($\Deta, \Dphi) = (0, 0)$ but the
shape and strength of these peaks differ markedly across models. The
strengths and widths of the peaks also evolve differently with
$\pt$. One notes, additionally, that the calculations for
$\PtwoCI$ exhibit quite noticeable differences with $\RtwoCI$: they feature narrower
peaks and different ordering in the
strengths predicted by the models. The RMS widths of these projections are plotted in
Fig.~\ref{fig:sevenPtWidthCi} and discussed
in more details in sec.~\ref{sec:widths}. It is clear at the outset, however, that
measurements of both
$\RtwoCI$ and $\PtwoCI$ in \pp\ collisions with
different $\pt$ ranges should in principle provide
significant constraints on the models and their underlying particle
production mechanisms.

Caution in the interpretation of the widths of the near-side peak of
the $\PtwoCI$ correlation functions is needed, however, because
of the complicated dependence of the correlation strength on the
distance to the centroid of the peak. One observes, in particular,
that the correlation strength of $\PtwoCI$ exhibits an
undershoot, in both $\Deta$ and $\Dphi$ projections,
in the $\pt$ range 2.0 - 5.0 $\gevc$, and a longer range oscillatory behavior in
projections of $\PtwoCI$ along $\Deta$ in the $\pt$ range 5.0 - 30.0 $\gevc$,
as expected from the angular ordering of particle $\pt$ discussed in
sec.~\ref{sec:definition}. While difficult to resolve, a hint for the existence of such
undershoot feature has already been reported in~\cite{AliceDptDptLongPaper}.
The presence of this undershoot stems from the explicit dependence of the correlator
on the particles' transverse momentum deviation from the mean, i.e.,
$\Delta \pt \Delta \pt$. At short angular distance
(both longitudinally and azimuthally), jet particles have momenta that
tend to exceed the mean $\pt$ and thus contribute positively, on
average, to the correlator. The presence of the undershoot
indicates that there is an angular range within
which the product $\Delta \pt \Delta \pt$ is negative on
average in PYTHIA events, but shifted in HERWIG
events. The shift observed in HERWIG events likely results from larger event-by-event
multiplicity fluctuations. At large angular
separation, both particles tend to have $\pt$ below the  $\la \pt\ra$ and thus contribute
positively to the $\Ptwo$
correlator. The peak and oscillatory behavior are thus determined by
the $\pt$ and angular ordering of the jet constituents. Given
PYTHIA and HERWIG produce particles using a different ordering, they are
expected and indeed observed to yield different shapes for the $\Ptwo$
correlation function. By contrast, the $\Rtwo$ correlation function
receives positive definite contributions from all particle pairs of a jet and is thus
not sensitive to the ordering of the particles of the pair but only
the overall width of the jet. Measurements of
$\Rtwo$ and $\Ptwo$ correlation functions in \pPb\ and \PbPb\ collisions
shall thus provide better discriminants of the parton splitting and hadronization
mechanisms at play in jet fragmentation as well as in the generation of the
underlying event.

The $\RtwoCI$ and $\PtwoCI$ correlations also exhibit
stark differences on the away-side, i.e., at
$\Dphi\sim \pi$. Inspection of the away-side of the $\Rtwo$
(Fig.~\ref{fig:cR2CI}) and $\Ptwo$ correlation functions
(Fig.~\ref{fig:cP2CI}), and their $\Dphi$ projections
(Fig.~\ref{fig:cDphiCI}) reveal the two correlators yield a rather
different response to the away-side jet. Indeed, the away-side jet
yields a relatively modest ridge-like structure at $\Dphi\sim\pi$ in
$\Rtwo$ correlation functions but produces a very large amplitude
away-side in $\Ptwo$. One also finds that PYTHIA and HERWIG produce
away-side ridges with different shapes and strengths as well as
quantitatively different $\pt$ dependence. Comparative measurements of
the $\Rtwo$ and $\Ptwo$ correlators in different particle momentum
ranges in \pp\ collisions should thus provide additional insight and
constraints on the hadronization mechanisms implemented in these models.

\subsubsection{Charge Dependent Correlations}
\label{sec:ChargeDependentCorrelations}

\begin{figure}[htb!]
 \begin{center} 
  \includegraphics[scale=0.8]{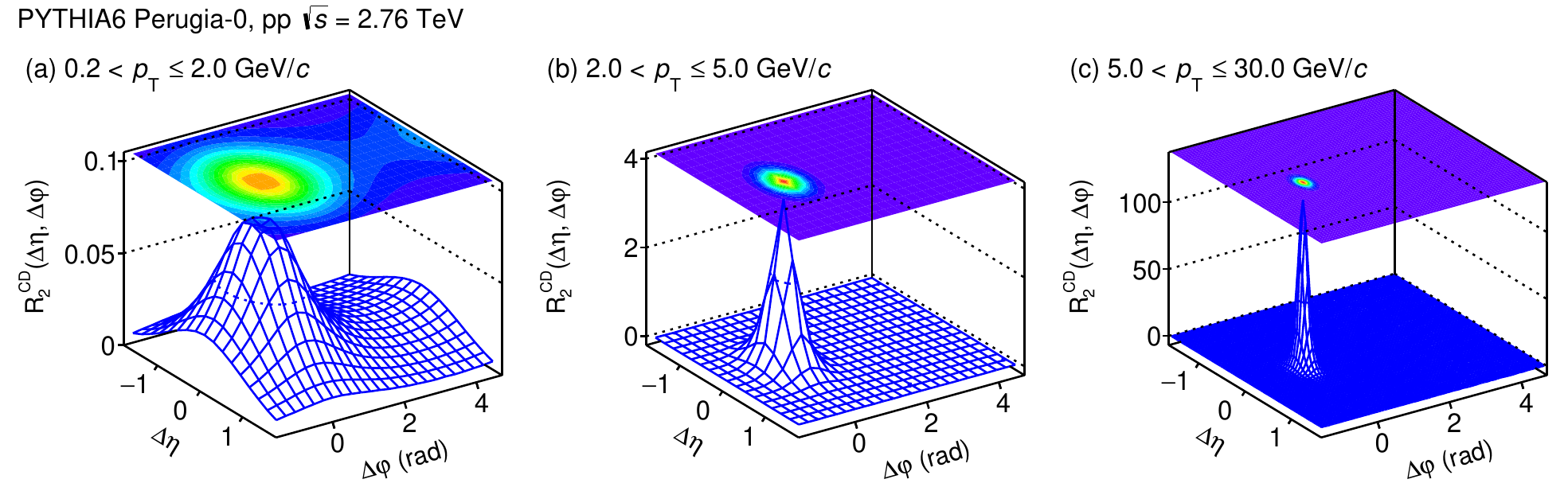}
  \includegraphics[scale=0.8]{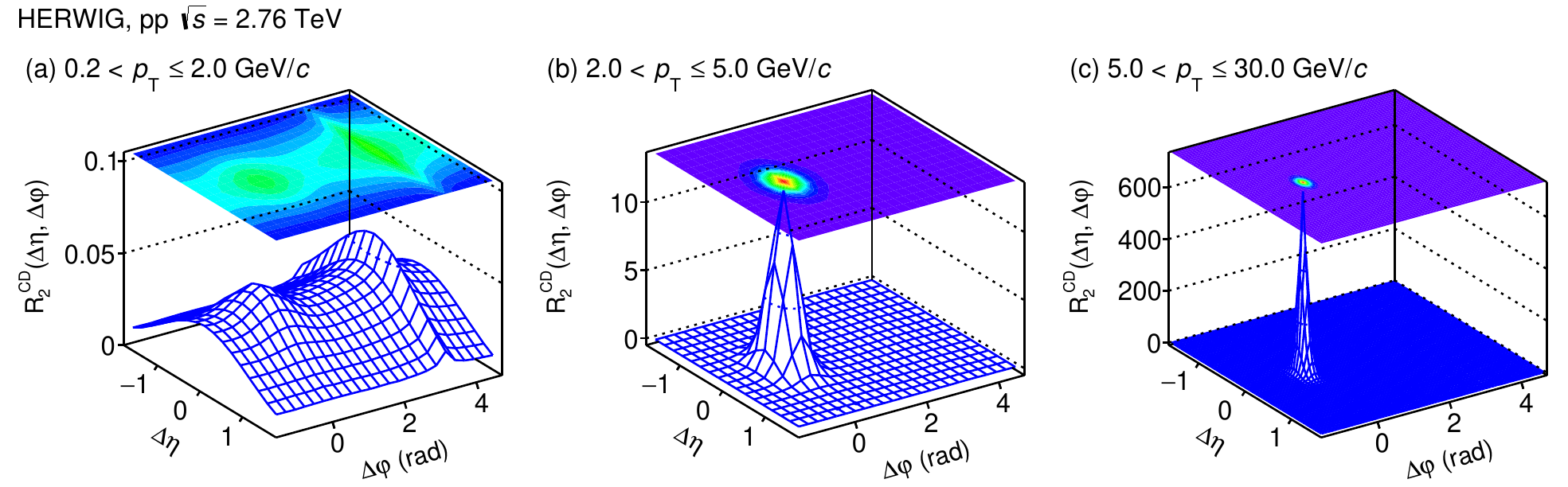}
  \caption{Correlation functions $\RtwoCD$ of charged 
   hadrons, in selected $\pt $ ranges, obtained with PYTHIA (top
   panel) and HERWIG (bottom panel) in \pp\ collisions at $\s$ = 2.76 TeV.}
  \label{fig:cR2CD}
 \end{center} 
\end{figure}

\begin{figure}[htb]
 \begin{center} 
  \includegraphics[scale=0.8]{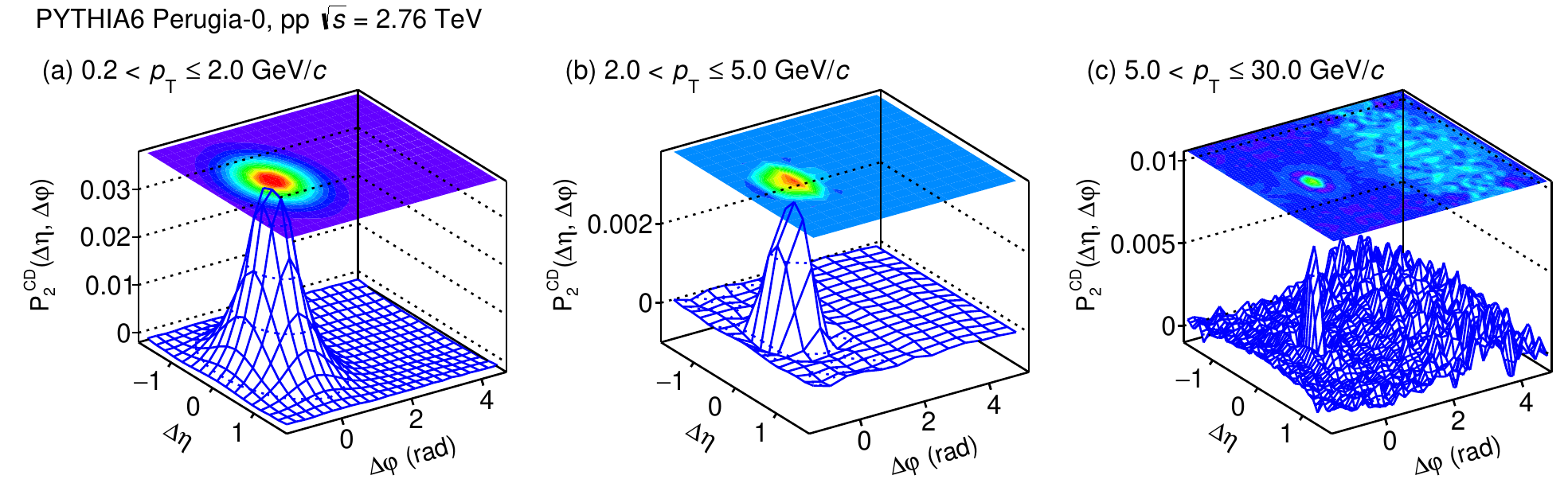}
  \includegraphics[scale=0.8]{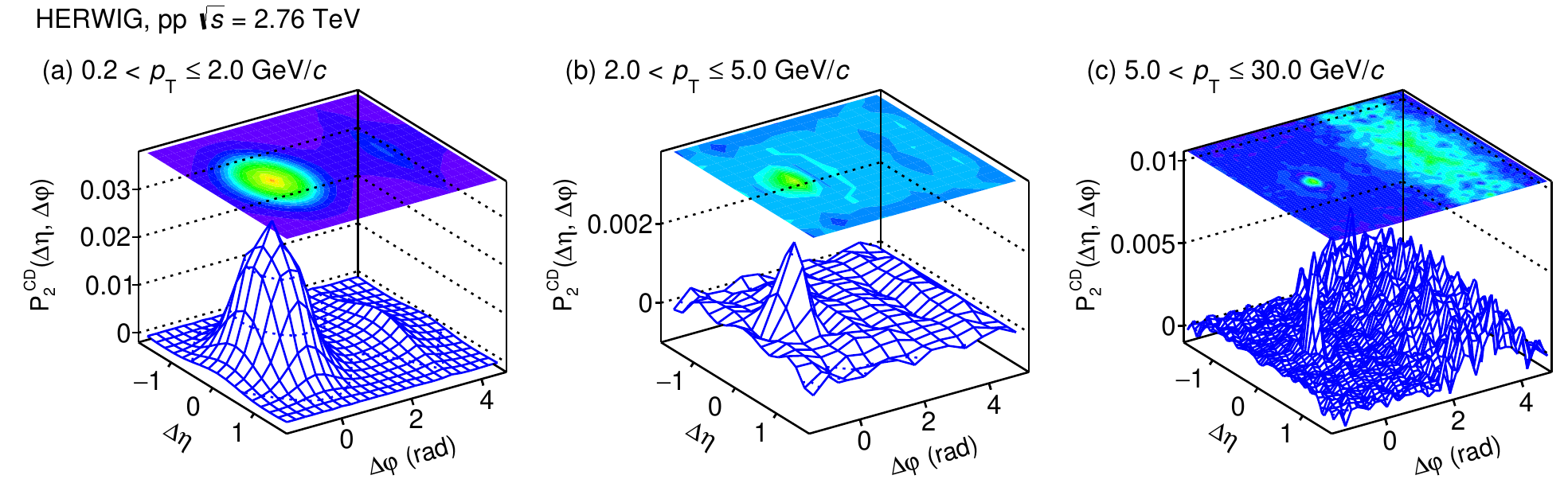}
  \caption{Correlation functions $\PtwoCD$ of charged
   hadrons, in selected $\pt $ ranges, obtained with PYTHIA
   (top panel) and HERWIG (bottom panel) in \pp\ collisions at $\s$ = 2.76 TeV.}
  \label{fig:cP2CD}
 \end{center} 
\end{figure}

We next shift our attention to the CD correlation functions presented
in Figs.~\ref{fig:cR2CD} and \ref{fig:cP2CD}.

\begin{figure}[htb!]
 \begin{center} 
  \includegraphics[scale=0.3]{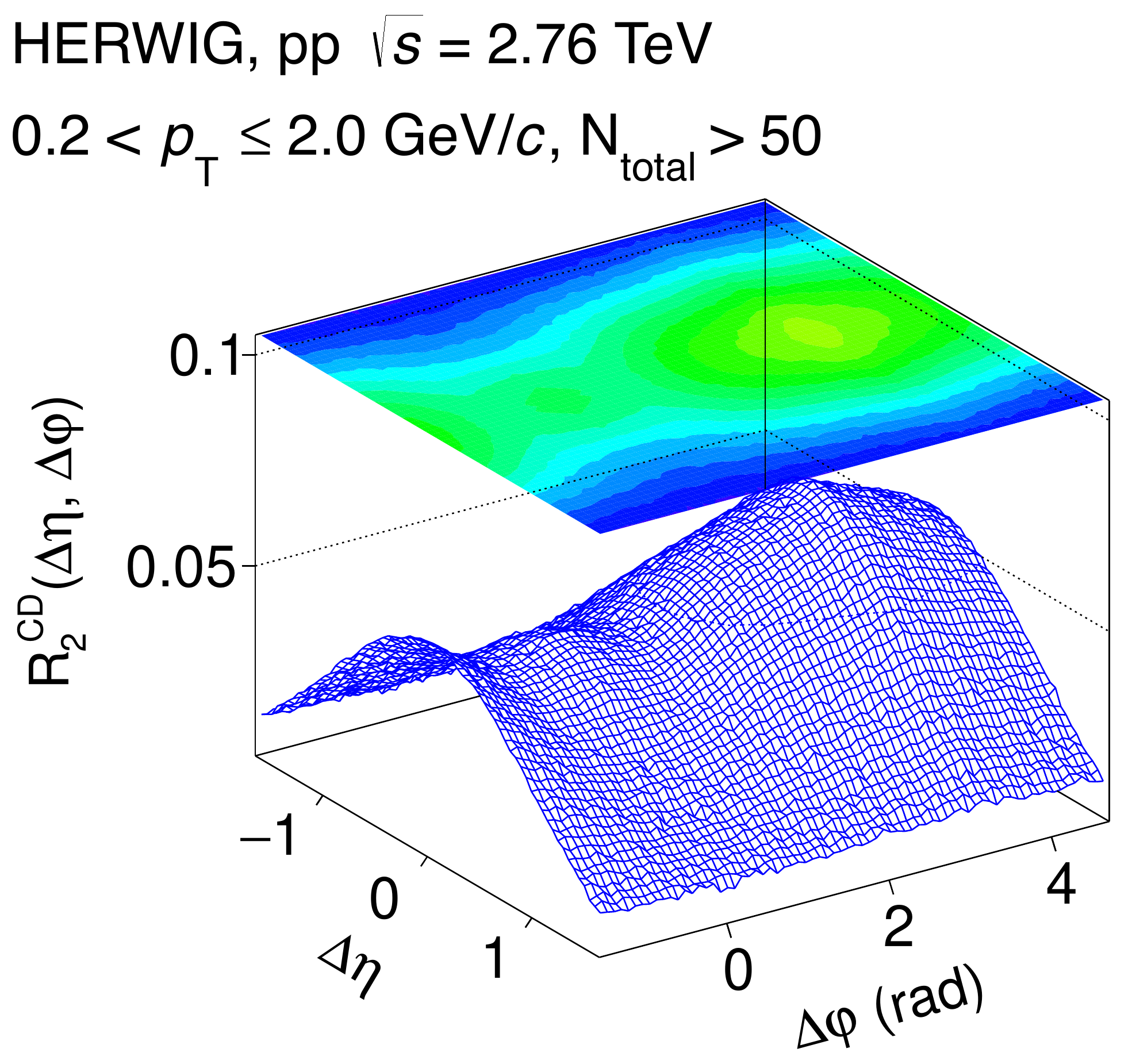}
  \caption{Correlation function $\RtwoCD$
   obtained with HERWIG for charged hadrons in \pp\ collisions at
   $\s$ = 2.76 TeV with a minimum multiplicity
   cut, $N_{\rm total} > 50$. }
  \label{fig:RtwoCDWcut}
 \end{center} 
\end{figure} 

\begin{figure}[ht!]
 \hspace*{-1.3cm}%
 \includegraphics[scale=0.4]{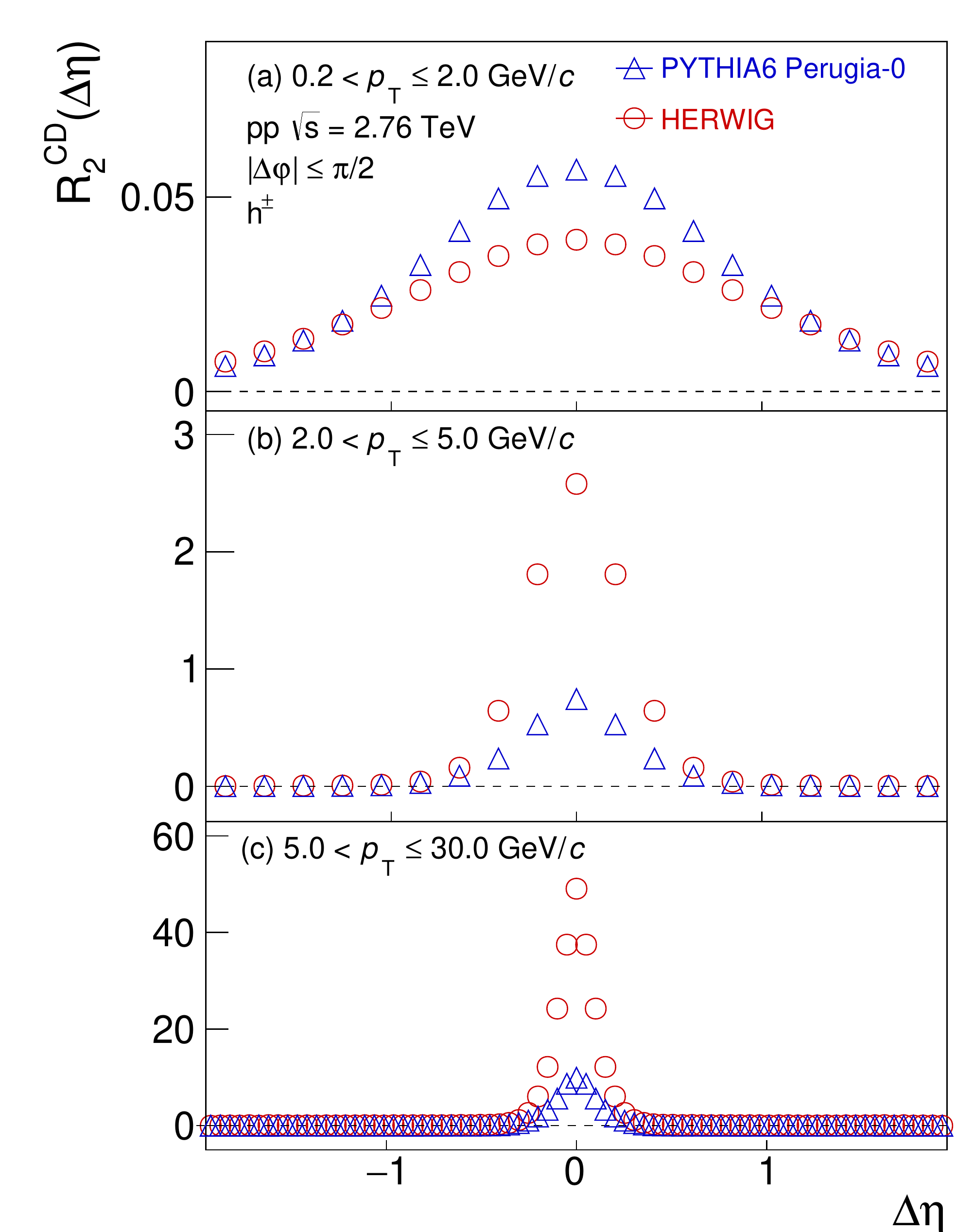}
 \includegraphics[scale=0.4]{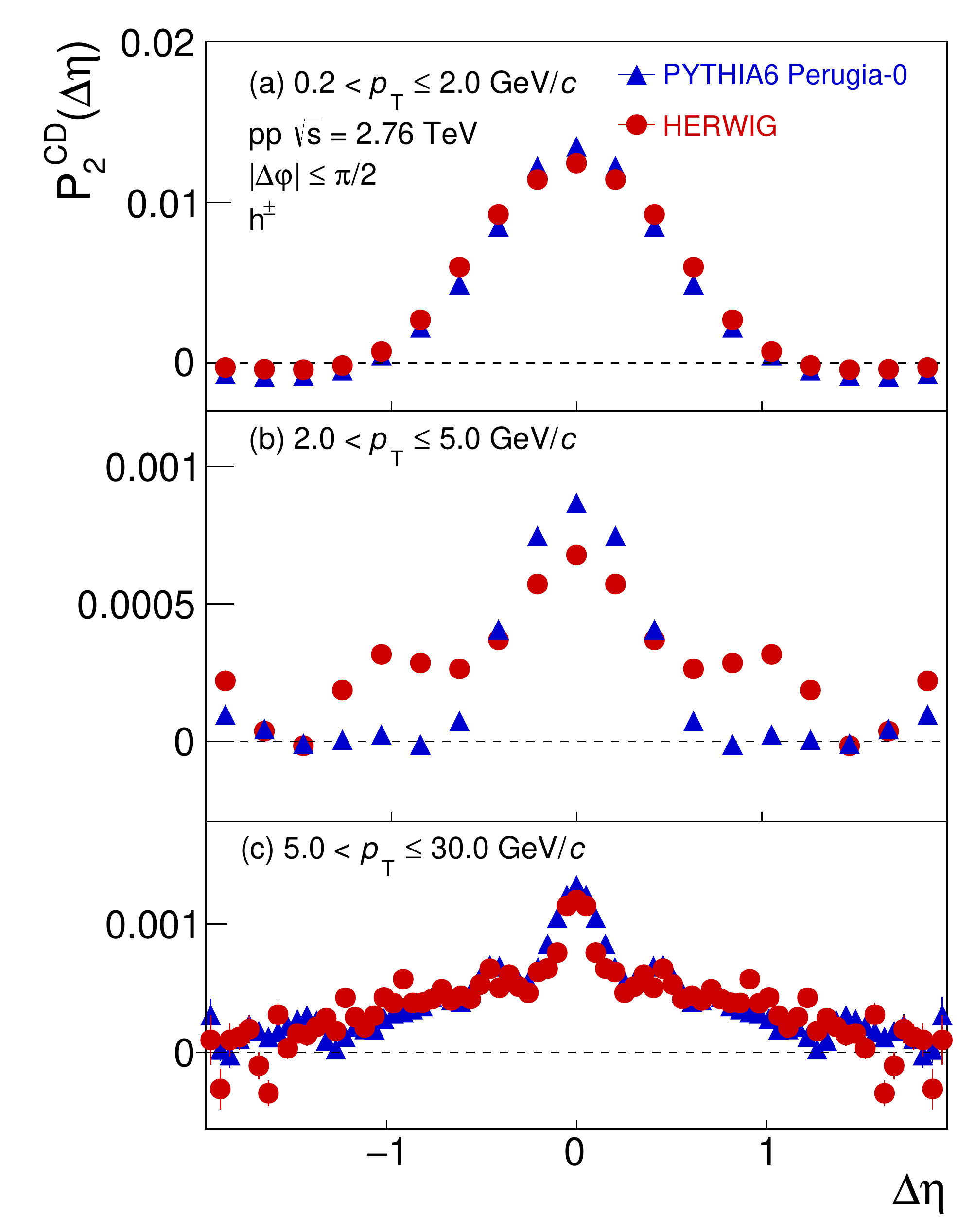}
 \caption{Projections onto $\Deta$ of $\RtwoCD$ (left column)
  and $\PtwoCD$ (right column) correlation functions of charged
  hadrons calculated in selected $\pt $ ranges with PYTHIA (blue)
  and HERWIG (red) in \pp\ collisions at $\s$ = 2.76 TeV. The
  projections are calculated as averages of the two-dimensional
  correlations in the range $|\Dphi| \leq \pi/2$.}
 \label{fig:cDetaCD}
\end{figure} 

These were obtained by
subtraction of the LS correlations from the US correlations according
to Eq.~(\ref{eq:CD}). As such, they emphasize the role of charge
conservation in particle production. Correlation functions $\Rtwo$
(and similarly charge balance functions) indeed provide signatures of the charged
particle pair production and transport in \pp\ and \AonA\ collisions.
For instance, at momenta in excess of 2.0 \gevc, as shown  in Fig.~\ref{fig:cR2CD}
(b,c), one observes the correlator $\Rtwo$ features an isolated peak centered at 
$(\Deta,\Dphi)=(0,0)$ resulting from the fact that correlated charged
particle production occurs almost exclusively within the confines of a
single jet.

\begin{figure}[ht!]
 \hspace*{-1.3cm}%
 \includegraphics[scale=0.4]{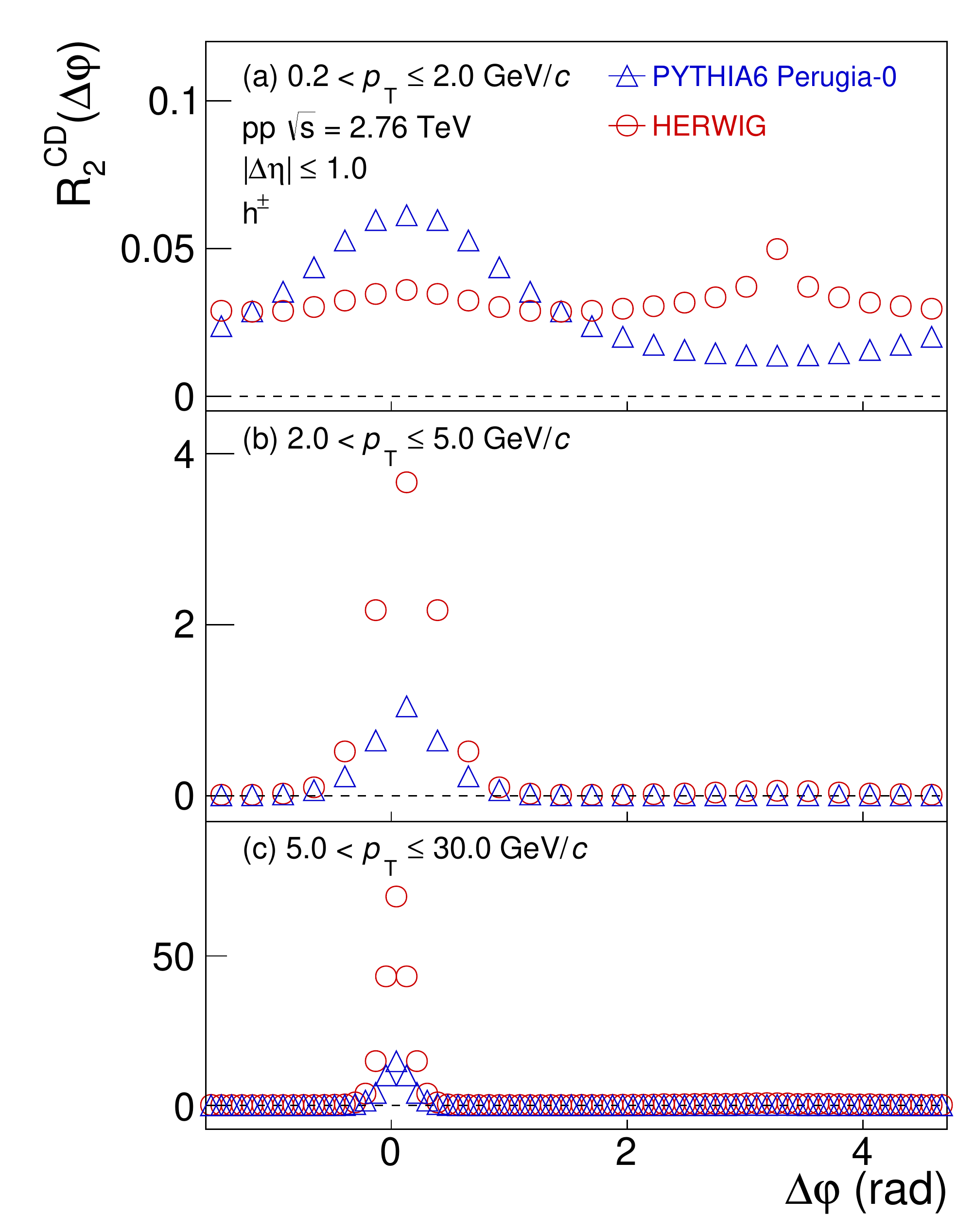}
 \includegraphics[scale=0.4]{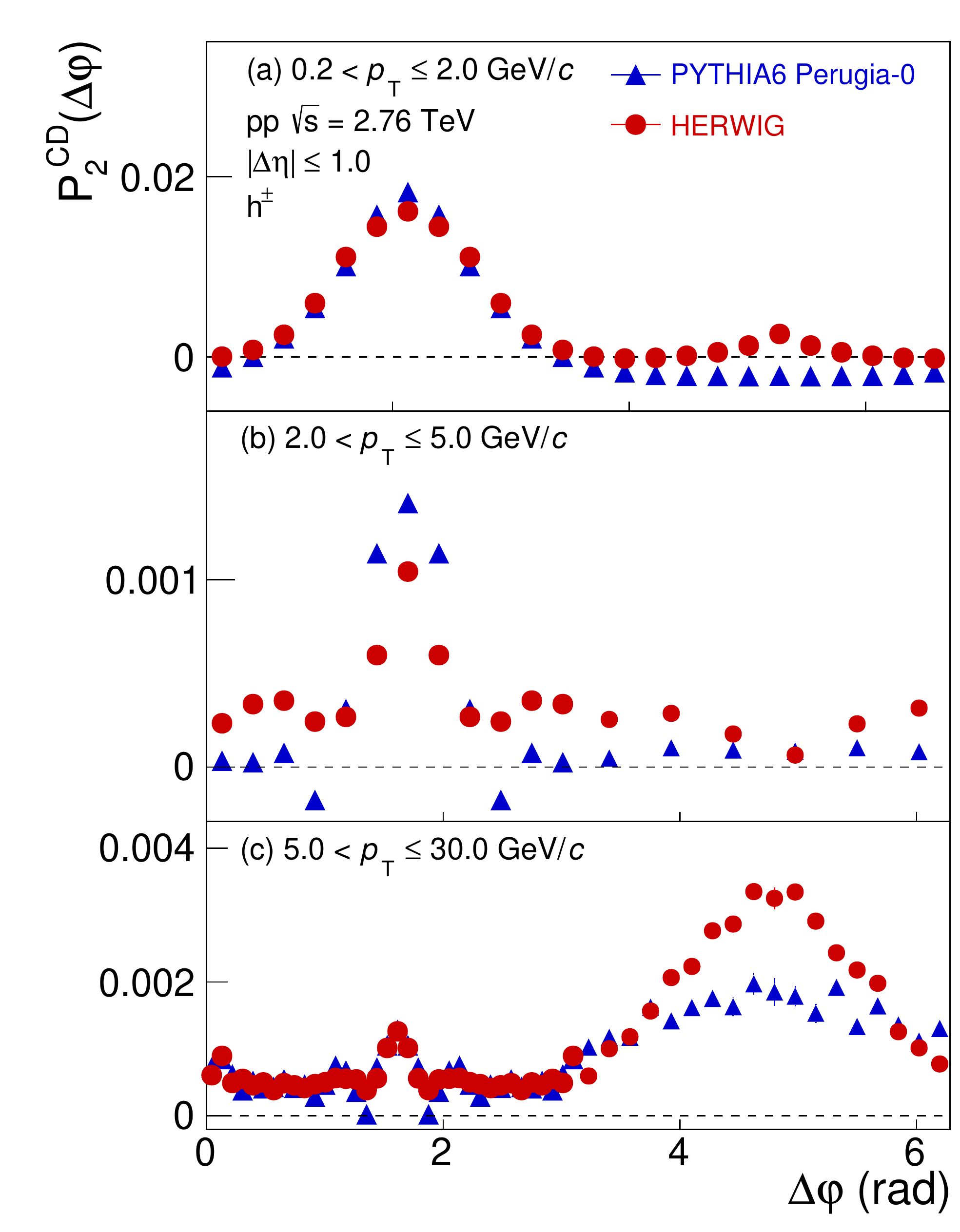}
 \caption{Projections onto $\Dphi$ of $\RtwoCD$ (left column)
  and $\PtwoCD$ (right column) correlation functions of
  charged hadrons calculated in selected $\pt $ ranges with PYTHIA
  (blue) and HERWIG (red) in \pp\ collisions at $\s$ = 2.76
  TeV. The $\Dphi$ projections are calculated as averages of the two-dimensional 
  correlations in the range $|\Deta| \leq 1.0$.}
 \label{fig:cDphiCD}
\end{figure}

The width of the peak decreases monotonically with increasing particle
momentum owing to the Schwinger mechanism and angular ordering discussed
in sec.~\ref{sec:definition}. At lower momenta, however,
correlated charged pair production may occur over a wider range of
angles, even back-to-back, as illustrated by the very sharp and narrow
away-side ridge predicted by HERWIG in the range
$0.2 \le \pt\le 2.0$ \gevc. The $\RtwoCD$ correlation
functions obtained with PYTHIA (top panels) and HERWIG (bottom panels) for particles
in the range $0.2 \le \pt \le 2.0$ \gevc\ indeed feature a more
complicated shape involving both a near-side peak and an away-side
structure. In this case, one notes that PYTHIA and HERWIG produce very
different predictions, owing most likely to their different implementation of the
underlying event. Additionally note, as exemplified in
Fig.~\ref{fig:cR2CD}, that the shape and strength of
$\RtwoCD$ exhibit a strong dependence on the produced hadron multiplicity.
Measurements of $\RtwoCD$ in \pp\ collisions for various
momentum and produced particle multiplicity ranges shall then provide
very useful constraints in the tuning of these models.

The PYTHIA and HERWIG predictions for $\PtwoCD$ correlation
functions, shown in Fig.~\ref{fig:cP2CD}, indicate this correlator is also of interest
to probe the internal structure of jets and the charge production
ordering. With PYTHIA, the near-side peak of the $\PtwoCD$
correlation functions is significantly narrower than its $\RtwoCD$ counterpart in the
lowest $\pt$ range considered, but somewhat
wider at higher $\pt$. By contrast, HERWIG's predictions of
$\PtwoCD$ features a somewhat narrower near-side peak in all
three momentum ranges. One also notes that the near-side predicted by
PYTHIA and HERWIG have different shapes, widths, and a somewhat
complicated dependence on the $\pt$ of the particles. One finds
additionally that the away-side of $\PtwoCD$ correlation
functions feature a large amplitude for high-$\pt$ particles. This is
in stark contrast to the $\RtwoCD$ correlation functions that
feature an almost flat away-side yield. Such a small away-side yield is expected in $\RtwoCD$ owing to the fact that particle
production above $\pt \ge 2$ \gevc\ is dominated by jet
fragmentation.
Given charge is conserved locally in the jet
fragmentation process, one can expect, on general grounds, that charge
correlations between jets, if any, are driven primarily by the charge
of the parton that initiate the jets. Quark jets may be charge
correlated but jets initiated by gluons should not be, at least to
first order. Measurements of the $\PtwoCD$ away-side strength
thus provide an additional tool to probe the nature of the jets
measured in \pp\ and \AonA\ collisions. Indeed, back-to-back gluon jets should
yield no contributions to the away-side of $\PtwoCD$
correlation functions but quark-quark jet pairs should have a finite
CD correlation. The measured away-side yield of $\PtwoCD$
correlation functions may thus provide a new tool to determine the
origin and nature of jets measured in elementary collisions.\\ 

We note that the $\RtwoCD$ correlator computed with HERWIG in
the range $0.2 < \pt \le 2.0$ \gevc, shown in  Fig.~\ref{fig:cR2CD},
features a narrow elongated structure at $\Delta\varphi=\pi$. This structure
likely corresponds to underlying event particle pairs emitted back-to-back in the
laboratory frame and are likely produced in  excess given such a structure is not
observed in data
reported by the ALICE collaboration~\cite{AliceDptDptLongPaper}. We find, however,
that contributions of such pairs to the correlator are suppressed in HERWIG events
featuring a total particle multiplicity  $N_{\rm total}$ > 50 in the fiducial acceptance, as
shown in Fig.~\ref{fig:RtwoCDWcut}. An experimental investigation of $\RtwoCD$ and $\PtwoCD$ correlation functions  as a function of \pp\ 
collision multiplicity is thus of obvious interest to elucidate the role 
and interplay of underlying events and multi-jet production in these collisions.

\subsection{Transverse momentum dependence of the width of the CI and CD near-side peaks}
\label{sec:widths}

We study the $\pt $ evolution of the RMS widths of the near-side peaks
of the $\Rtwo$ and $\Ptwo$ correlation functions obtained with
PYTHIA and HERWIG. The $\Deta$ and $\Dphi$ RMS widths are calculated
according to 
\be
\sigma_{\Delta \eta} &=& 
\left(
 \frac{\sum_{i,j} [\rm{O}(\Deta_{i},\Dphi_{j}) -O_{\rm offset}] \Delta \eta_{i}^{2}}
 {\sum_{i,j} \rm{O}({\Deta}_{i}, {\Dphi}_{j})} 
\right)^{1/2}, \\ 
\sigma_{\Dphi} &=& 
\left(
 \frac{\sum_{i,j} [\rm{O}(\Deta_{i},\Dphi_{j}) -O_{\rm offset}] \Dphi_{i}^{2}}
 {\sum_{i,j} \rm{O}({\Deta}_{i}, {\Dphi}_{j})} 
\right)^{1/2},
\ee
where $\rm{O}({\Deta}_{i}, {\Dphi}_{j})$ represent the strength of the
correlation functions in bins $\Deta_i$ and $\Dphi_j$ and the sums on
$\Deta_i$ covers the $|\Deta| \le 1.0$ acceptance of the simulation,
whereas the sums on $\Dphi_i$ are limited to exclude the away-side ridge.
Offsets are used to suppress negative correlation values and eliminate trivial width
values determined by the breadth of the acceptance. They are
obtained by taking the average of three bins at the edge of the acceptance, i.e., at $|\Deta|$ = 2.0 for $\Deta$ projections, and at the minimum of the correlations, near
$\Dphi$ = -$\pi/2$, for $\Dphi$ projections. The three-bin average
technique is also used for calculating  offsets whenever undershoots are present 
in $\Ptwo$ correlators.

\begin{figure}[ht!]
 \hspace*{-1.3cm}%
 \includegraphics[scale=0.80]{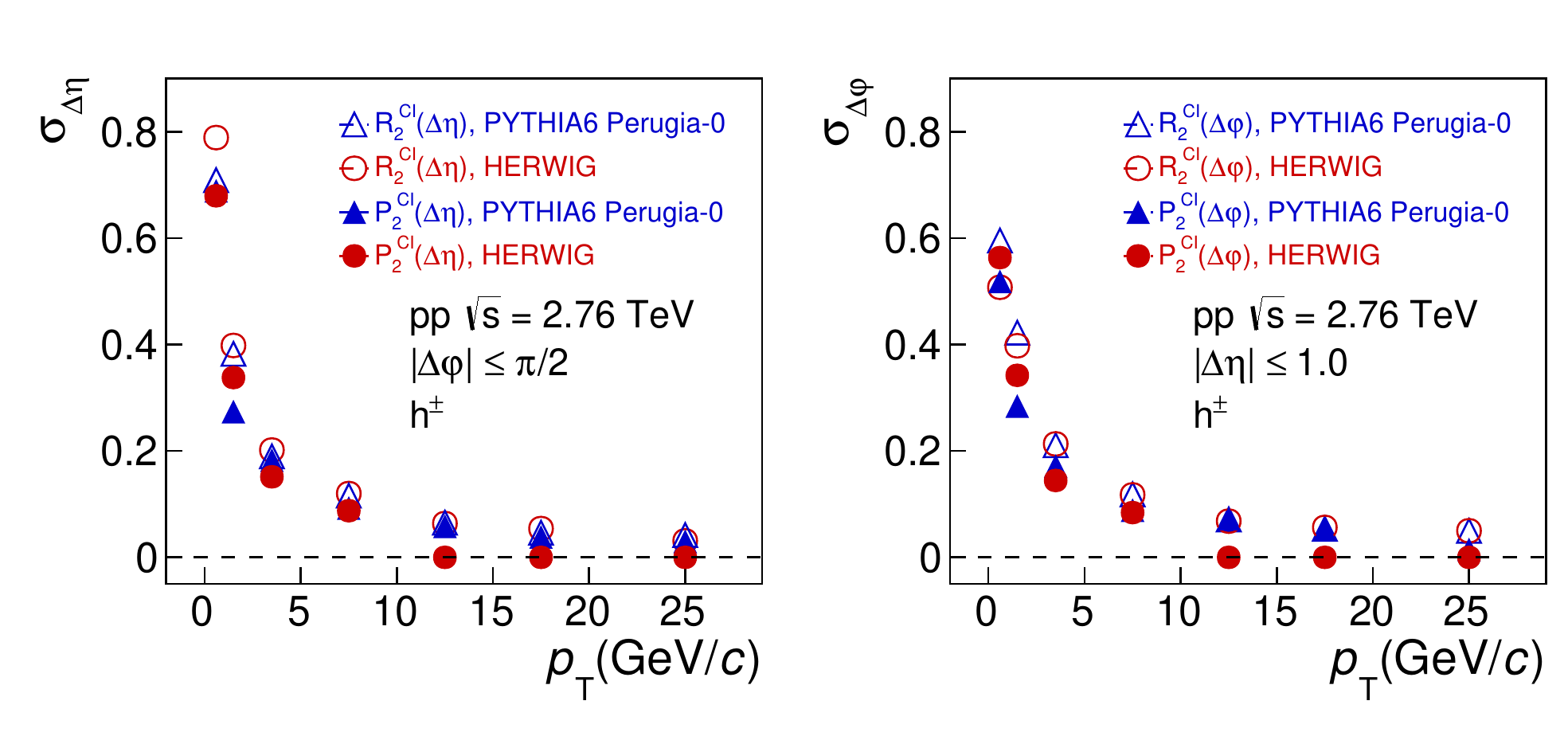}
 \caption{Width of the near-side peak of CI correlation functions
  along $\Deta$ (left panel), in the range $|\Dphi| \leq \pi/2$, and along 
  $\Dphi$ (right panel), in the range $|\Deta| \leq 1.0$, as
  function of the momentum of the particles.}
 \label{fig:sevenPtWidthCi}
\end{figure}
\begin{figure}[ht!]
 \hspace*{-1.3cm}%
 \includegraphics[scale=0.80]{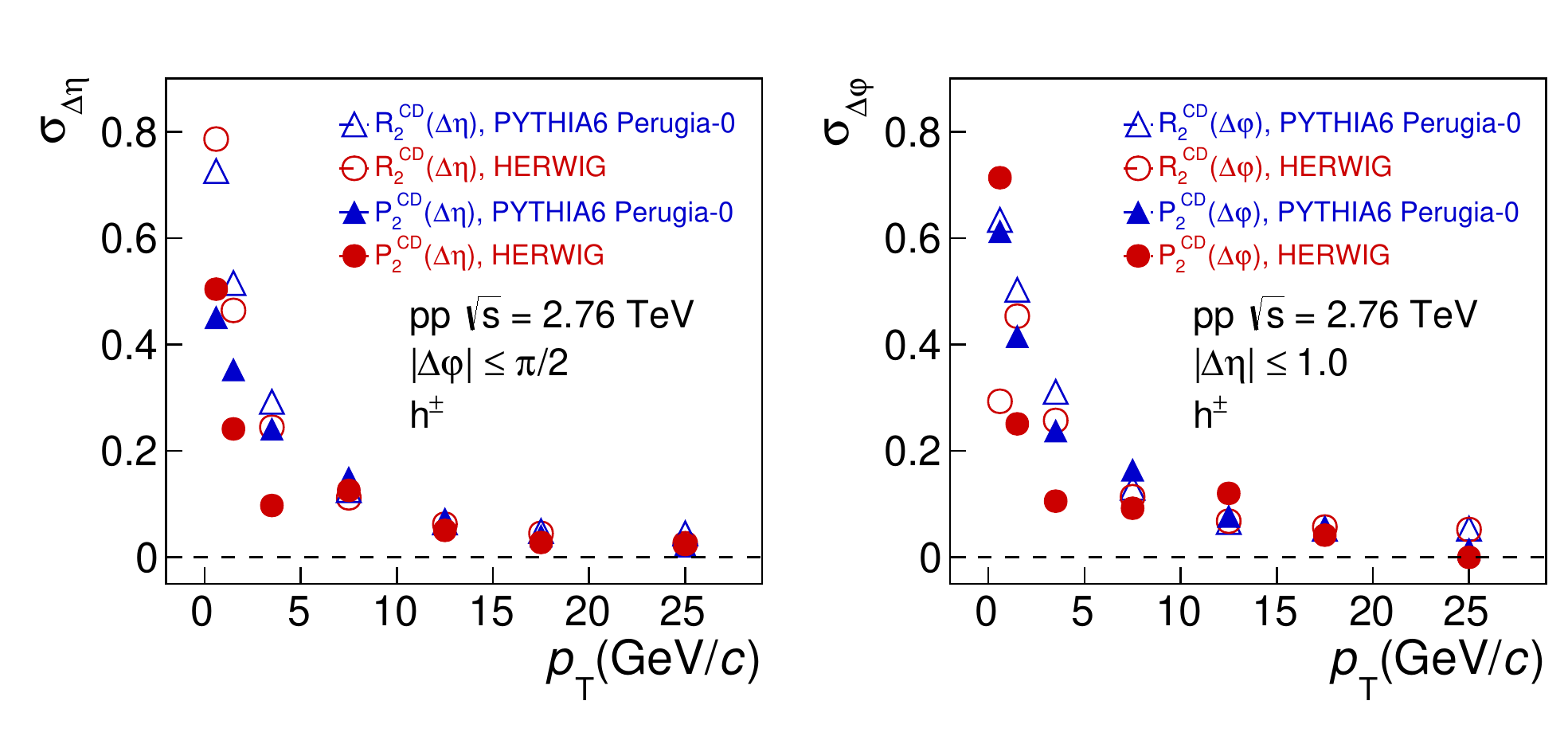}
 \caption{Width of the near-side peak of CD correlation functions
  along $\Deta$ (left panel), in the range $|\Dphi| \leq \pi/2$, and along 
  $\Dphi$ (right panel), in the range $|\Deta| \leq 1.0$, as
  function of the momentum of the particles.}
 \label{fig:sevenPtWidthCd}
\end{figure}
\begin{figure}[ht!]
 \hspace*{-1.3cm}%
 \includegraphics[scale=0.45]{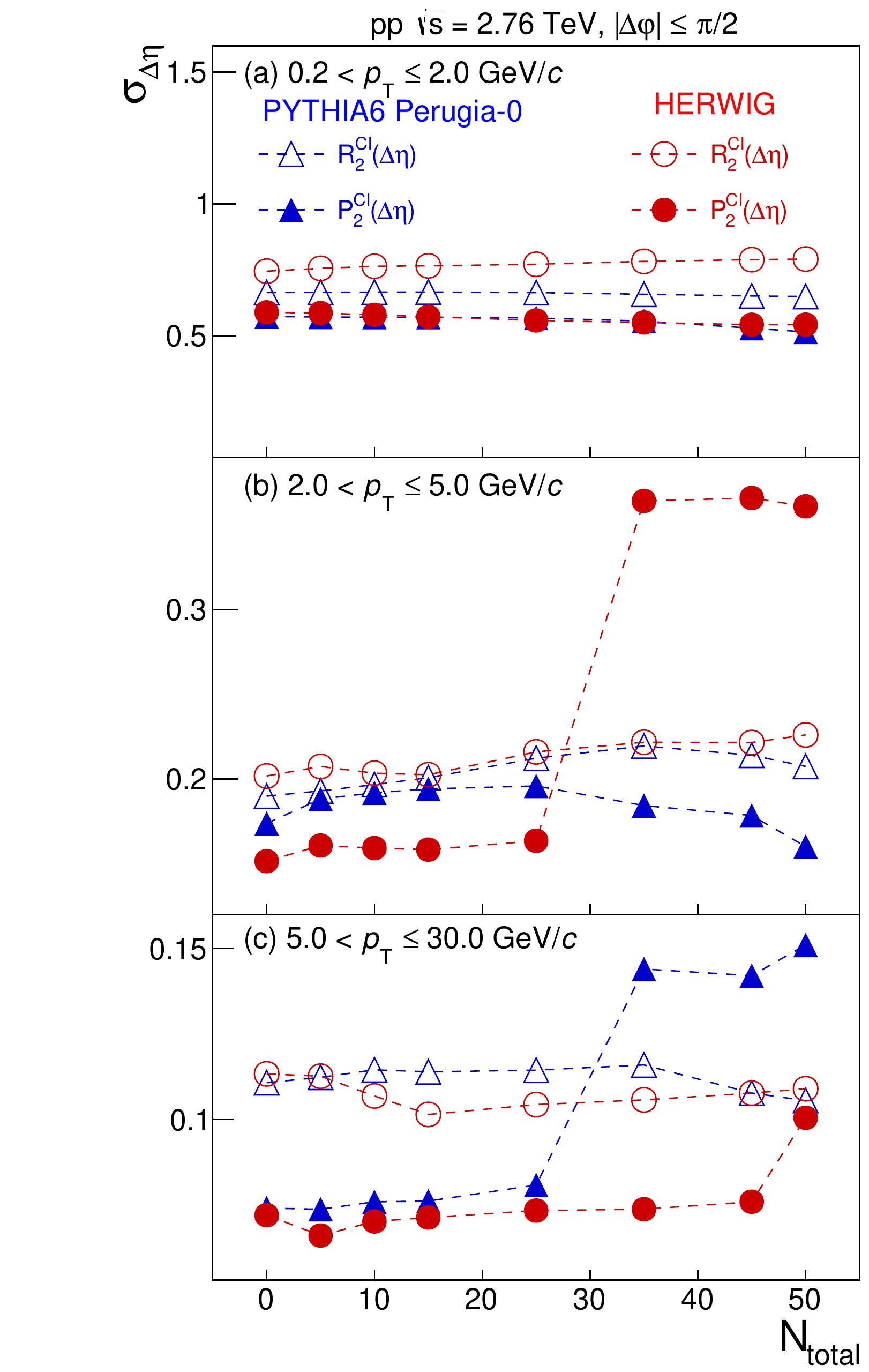}
 \includegraphics[scale=0.45]{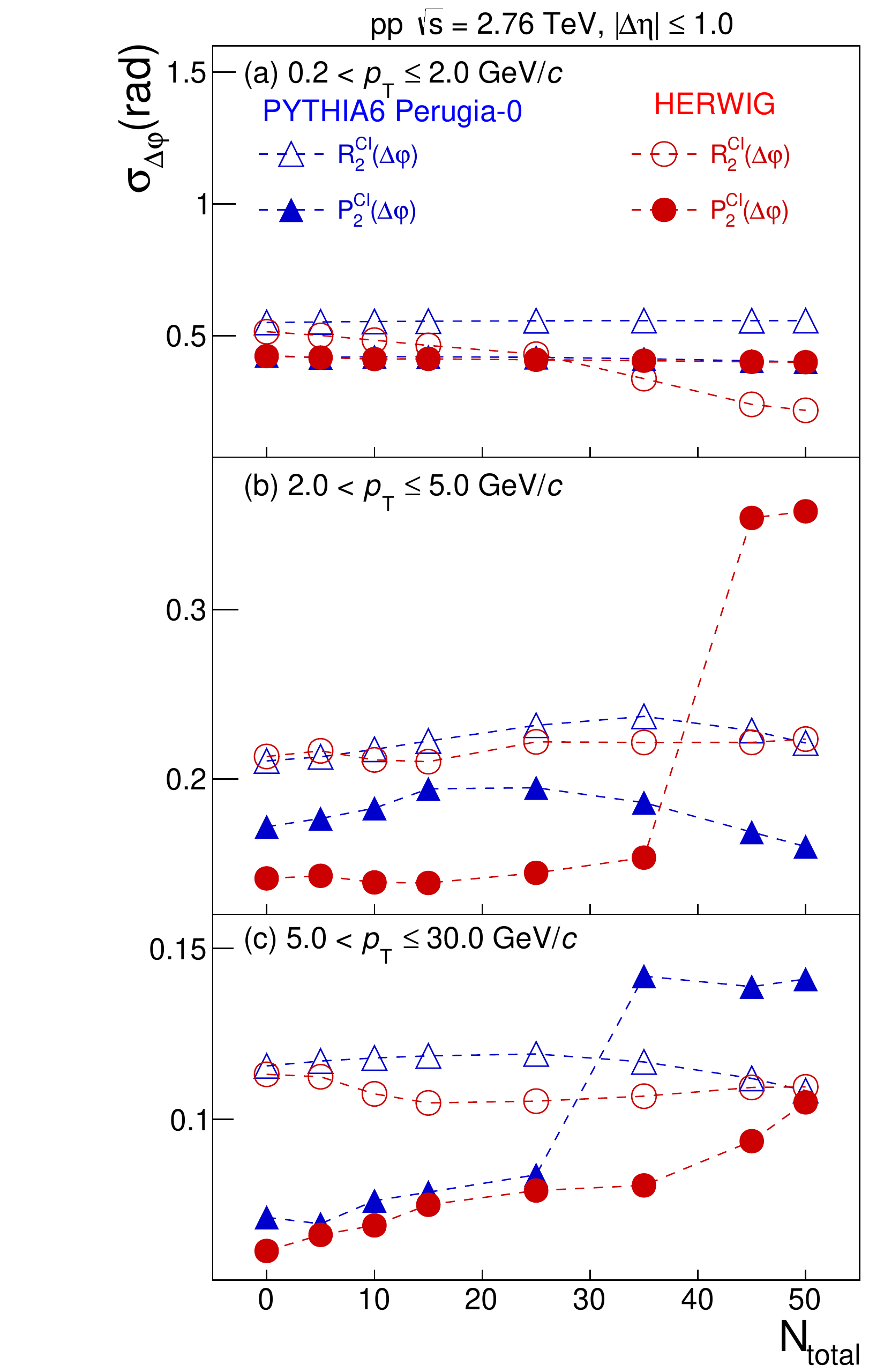}
 \caption{Width of the near-side peak of CI correlation functions
  along $\Deta$ (left panel), in the range $|\Dphi| \leq \pi/2$,
  and along $\Dphi$ (right panel), in the range $|\Deta| \leq 1.0$. Dotted lines are drawn to guide the eye.}
 \label{fig:ci3Pt8NchCutWidth}
\end{figure}
\begin{figure}[ht!]
 \hspace*{-1.3cm}%
 \includegraphics[scale=0.45]{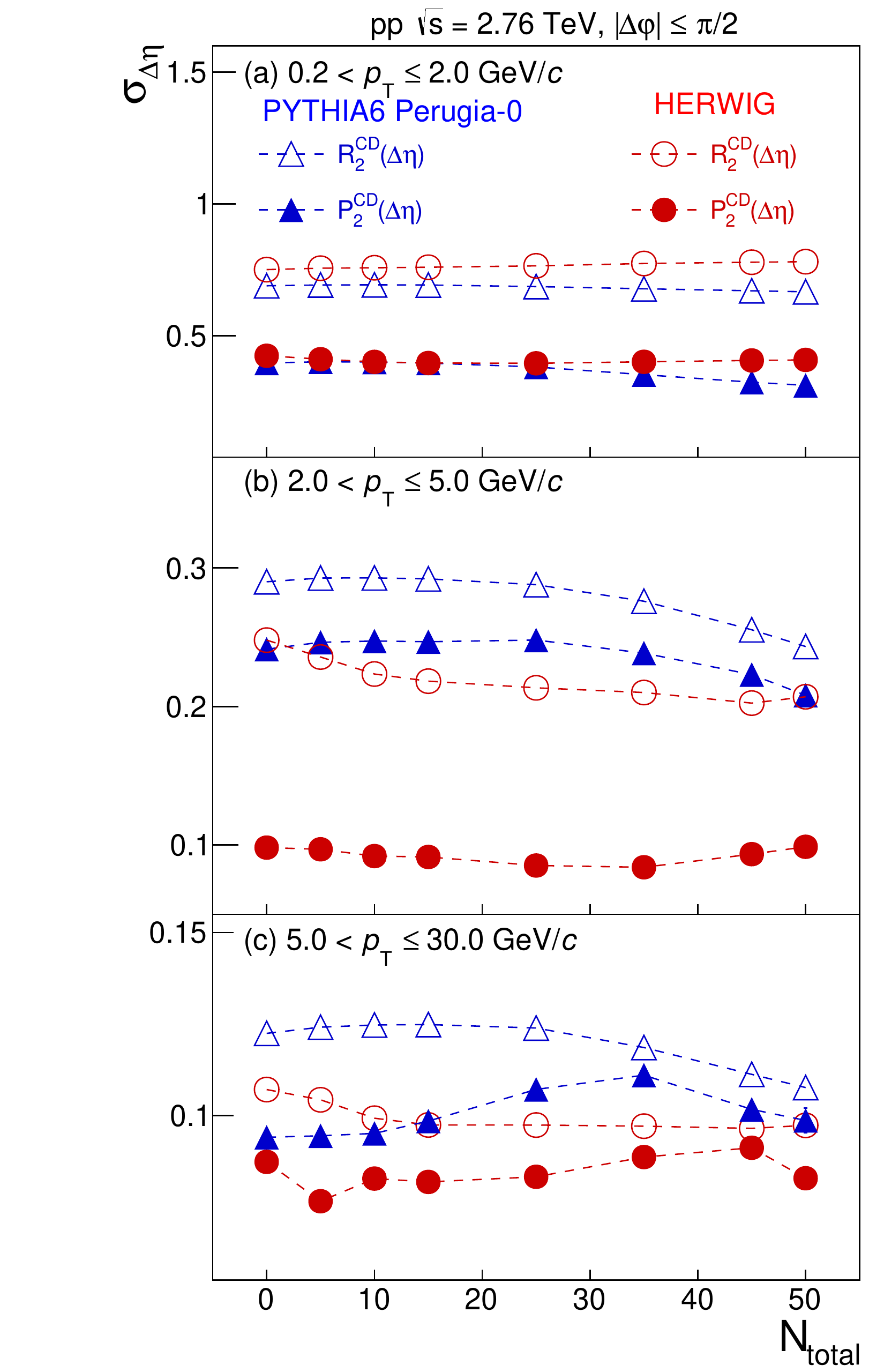}
 \includegraphics[scale=0.45]{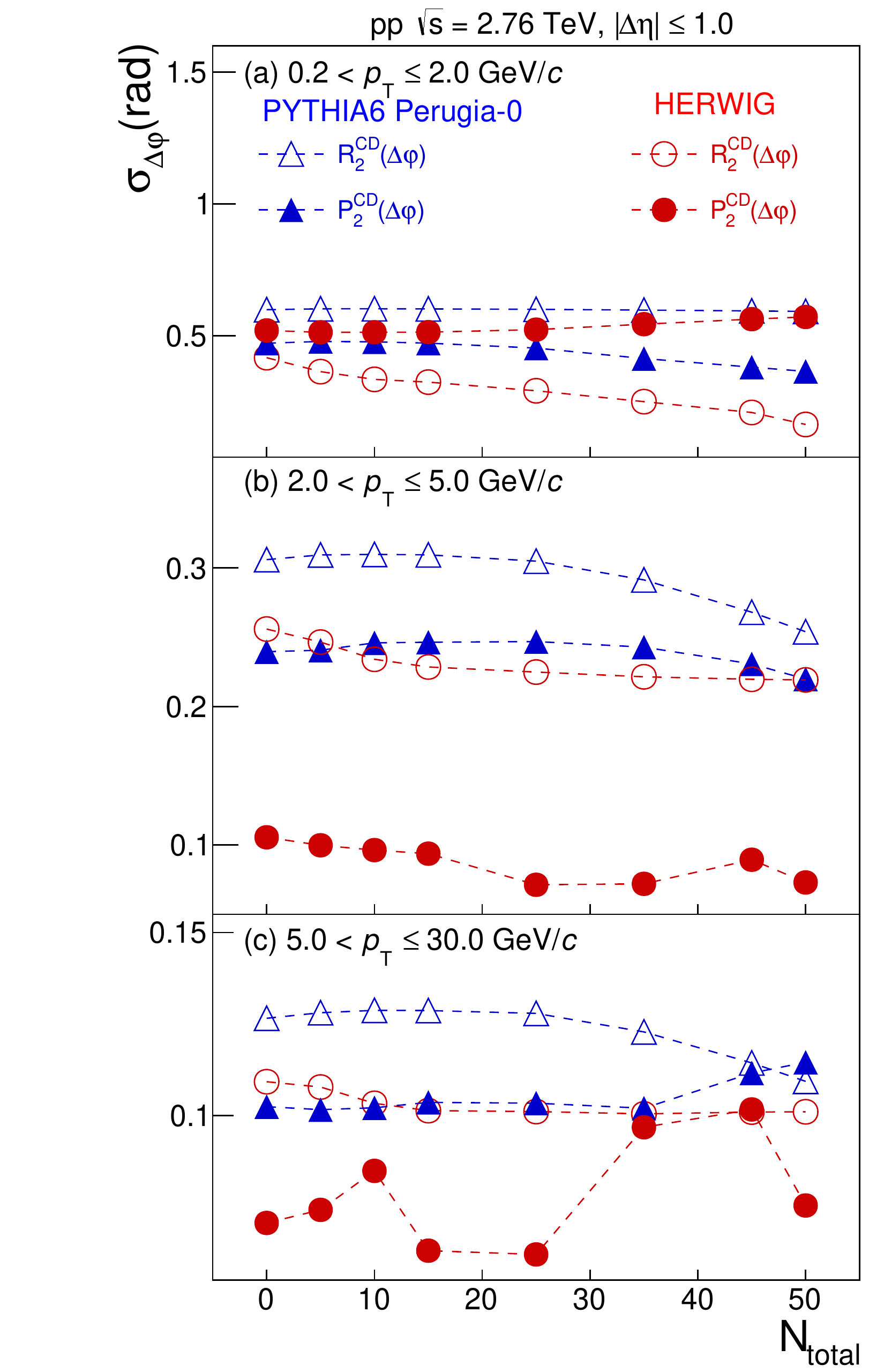}
 \caption{Width of the near-side peak of CD correlation functions
  along $\Deta$ (left panel), in the range $|\Dphi| \leq \pi/2$, and along 
  $\Dphi$ (right panel), in the range $|\Deta| \leq 1.0$. Dotted lines are drawn to guide the eye.}
 \label{fig:cd3Pt8NchCutWidth}
\end{figure}
\begin{figure}[ht!]
 \hspace*{-1.3cm}%
 \includegraphics[scale=0.35]{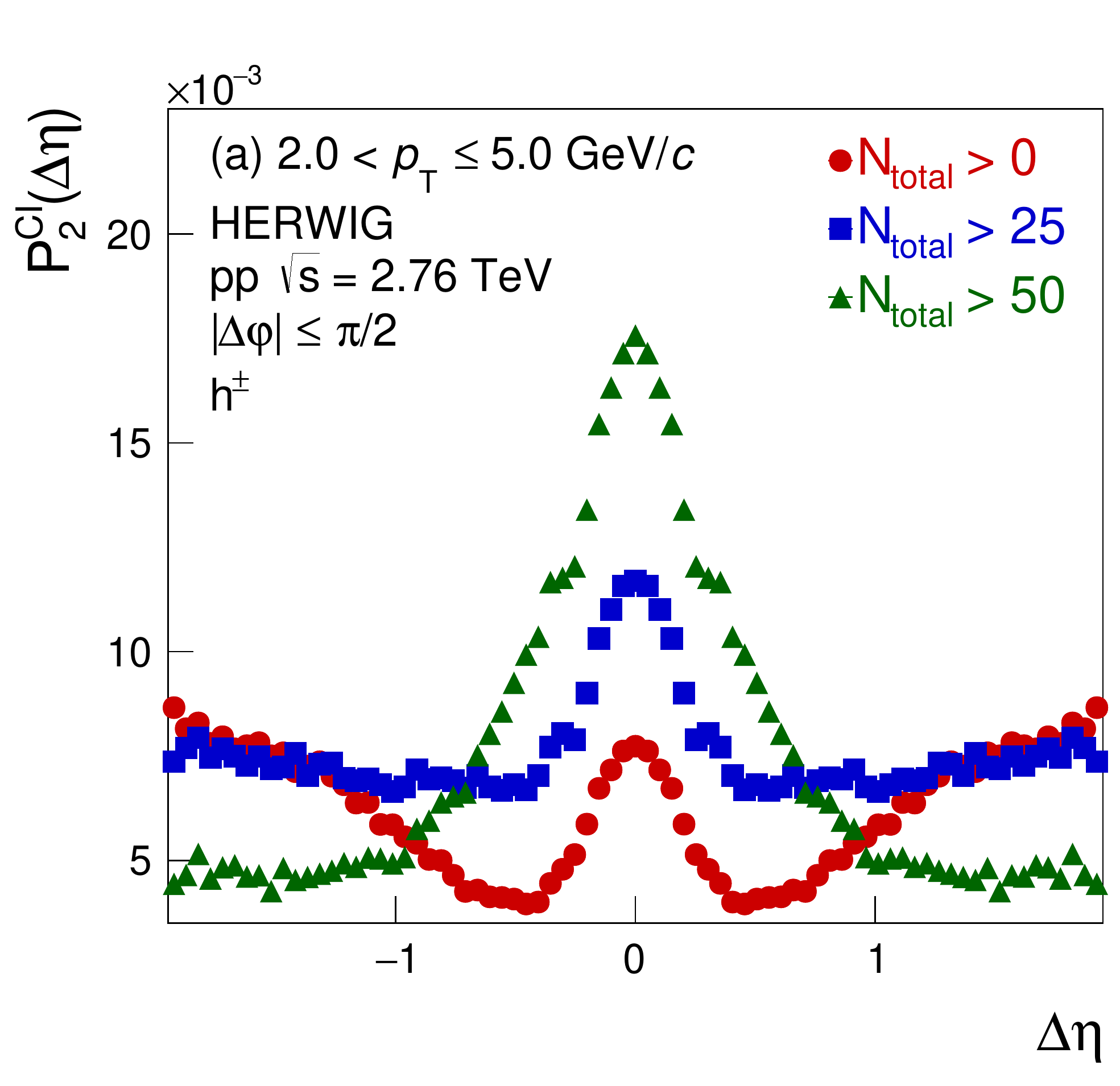}
 \includegraphics[scale=0.35]{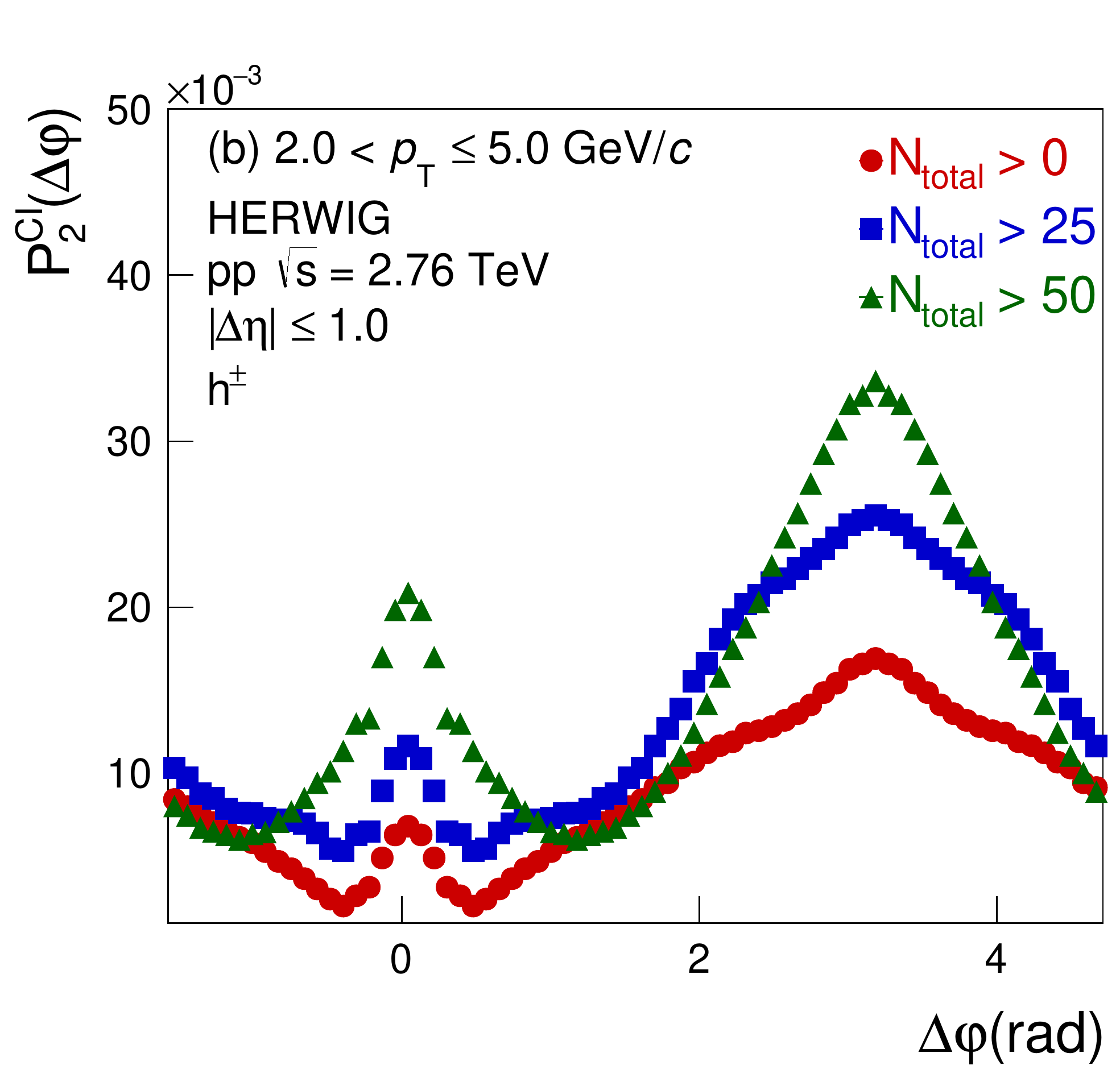}
 \caption{Projections onto $\Deta$ (left
  panel) and $\Dphi$ (right panel) of the $\PtwoCI$ correlation functions
  calculated with HERWIG for $h^{\pm}$ in the $\pt$ range 2.0 - 5.0 $\gevc$ in
  \pp\ collisions at $\s$ = 2.76 TeV for 3 different $N_{\rm total}$ cuts. The $\Deta$ and $\Dphi$ projections are calculated as averages of the two-dimensional correlations in the
 ranges $|\Dphi| \leq \pi/2$ and $|\Deta| \leq 1.0$,
respectively.}
 \label{fig:midPtDphiP2Herwig}
\end{figure} 

Figures~\ref{fig:sevenPtWidthCi} and \ref{fig:sevenPtWidthCd} present plots of the
evolution of the $\Deta$ and $\Dphi$ widths of the near-side peak of CI and CD
correlators  as a function of $\pt$. Overall,
one finds  the widths decrease with rising $\pt$. However, widths obtained with PYTHIA
exhibit a smooth and monotonic behavior with increasing particle $\pt$ whereas widths
obtained from HERWIG exhibit a more complicated $\pt$ dependence.
This study reveals an interesting case where the $\Ptwo$ width is
 broader than that of $\Rtwo$ in some $\pt$ ranges, in stark contrast
 with the results reported in~\cite{AliceDptDptLongPaper}.
 
In order to further understand the structures observed in $\Rtwo$ and $\Ptwo$
correlation functions presented in Fig.~\ref{fig:cR2CI} - \ref{fig:cP2CD}, we study, in
Figs.~\ref{fig:ci3Pt8NchCutWidth} and \ref{fig:cd3Pt8NchCutWidth}, the evolution of
the near-side peak of the correlators as a function of the  total particle multiplicity
$N_{\rm total}$. We find that the longitudinal and azimuthal widths,
$\sigma_{\Delta\eta}$ and $\sigma_{\Delta\varphi}$, of the $\RtwoCI$, $\RtwoCD$,
and $\PtwoCD$ correlators are slowly varying functions  of $N_{\rm total}$,
with largest dependence observed for the width $\sigma_{\Delta\varphi}$
predicted by HERWIG for particles in the range $0.2 < \pt \le  2.0$ \gevc. The
widths of the $\PtwoCI$ correlator, on the other hand, exhibit
a more complicated dependence on  $N_{\rm total}$. One observes, indeed,
that the widths extracted both from PYTHIA and HERWIG exhibit
a discontinuity near or above $N_{\rm total}=30$, thereby signaling a drastic
change in the shape of these correlation functions between low and high multiplicity
events. The shape dependence on $N_{\rm total}$ is illustrated in
Fig.~\ref{fig:midPtDphiP2Herwig}. Events of low multiplicity feature $\PtwoCI$ correlator 
with a clear undershoot structure, yielding  narrow widths in both  the longitudinal and
azimuthal directions. As argued above, the undershoot
structure is associated with
the production of pairs featuring $\Delta \pt \Delta \pt <0$ but multiplicity fluctuations
shift the correlator, globally, to positive values. The number
of such pair combinations is manifestly reduced, however, for collisions
with large  $N_{\rm total}$. These consequently do not feature an undershoot behavior
and thus produce a broad near-side peak. This behavior likely stems
from the fact that  high-multiplicity events favor gluon jets. These are less collimated
than quark jets and feature softer particles on average~\cite{quarkGluonJetComp}.
Evidently, such variations are not possible with the $\RtwoCI$ correlators. We thus
conclude that the $\PtwoCI$ correlator constitutes a more discriminating probe of the
correlation structure of jets and their underlying events than the $\RtwoCI$ correlator.

\subsection{Identified charged hadron correlations} 
\label{sec:ResultsPID}
\begin{figure}[htb] 
 \includegraphics[scale=0.8]{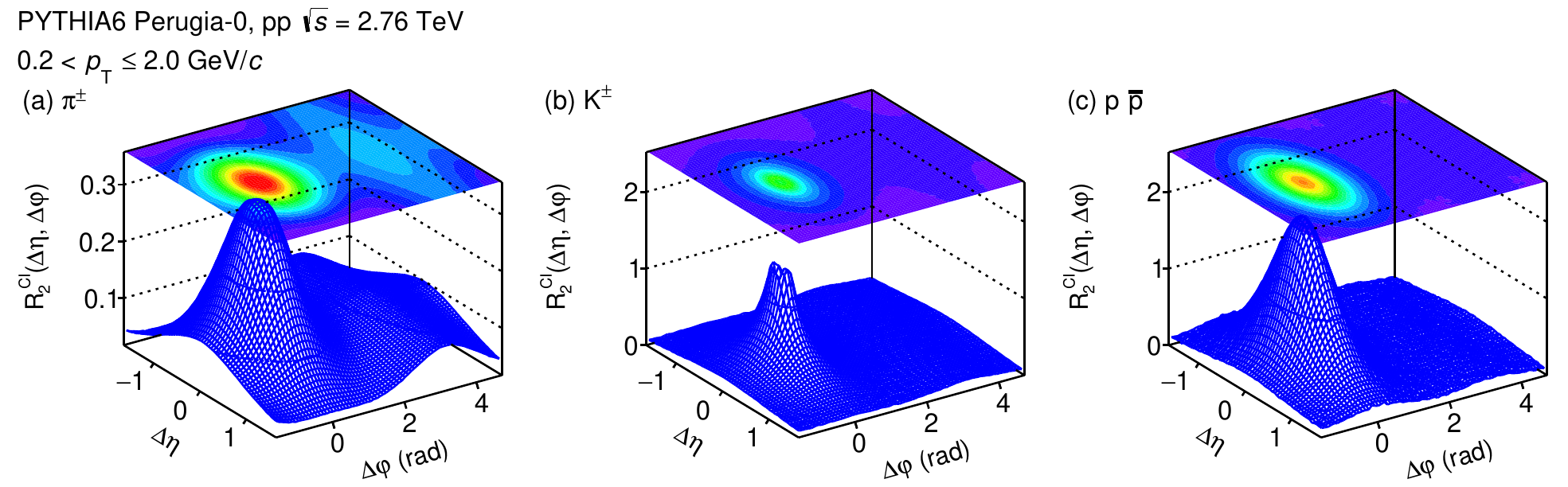}
 \caption{Correlation functions $\RtwoCI$ of $\pi^\pm$,
  $K^\pm$ and $p\bar{p}$, within $|\eta| < 1.0$
  and 0.2 < $\pt $ $\leq$ 2.0 $\gevc$, obtained with PYTHIA in \pp\
  collisions at $\s$ = 2.76 TeV.  }
 \label{fig:cR2CIidentified} 
\end{figure}

\begin{figure}[htb]
 \includegraphics[scale=0.8]{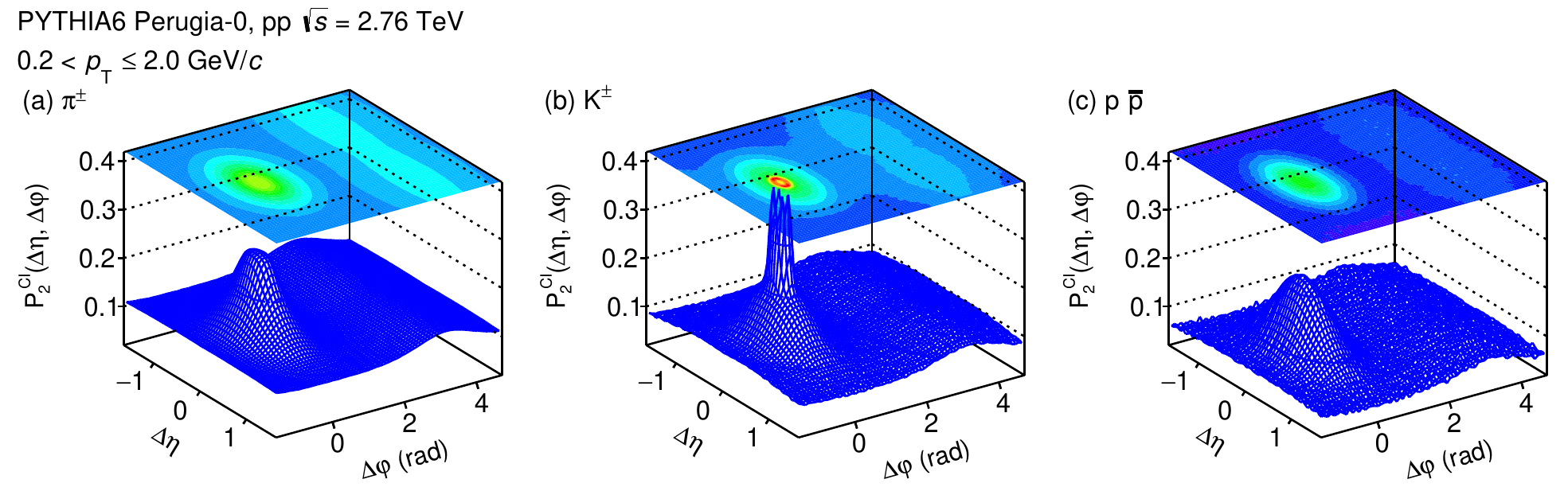}
 \caption{Correlation functions $\PtwoCI$ of $\pi^\pm$,
  $K^\pm$ and $p\bar{p}$, within $|\eta| < 1.0$
  and 0.2 < $\pt $ $\leq$ 2.0 $\gevc$, obtained with PYTHIA in \pp\
  collisions at $\s$ = 2.76 TeV.}
 \label{fig:cP2CIidentified} 
\end{figure}

Experimental studies at the ISR, FNAL, and the LHC have shown that 
jet fragmentation functions of identified species vary appreciably
between mesons and baryons as well as with their quark
content (\cite{PhysRevD.98.030001} and references therein). Unfortunately, measuring
the fragmentation functions of identified hadron species within jets is a statistically
onerous and difficult task, specially for high-$\it{z}$ particles in high-energy jets.
Measuring the strength and shape of $\RtwoCI$, 
$\PtwoCI$, $\RtwoCD$, and $\PtwoCD$ correlation
functions of identified high-$\pt$ hadrons, however, may provide an
invaluable proxy to such studies. We proceed to substantiate this
hypothesis by studying the shape and strength of identified hadron
correlation functions based on predictions by the PYTHIA and HERWIG models.
Figures \ref{fig:cR2CIidentified} and \ref{fig:cP2CIidentified}
respectively display $\RtwoCI$ and $\PtwoCI$
correlation functions calculated with PYTHIA for $\pi^\pm$, $K^\pm$ and $p\bar{p}$,
in the range 0.2 < $\pt $ $\leq$ 2.0 $\gevc$.

\begin{figure}[htb] 
 \includegraphics[scale=0.8]{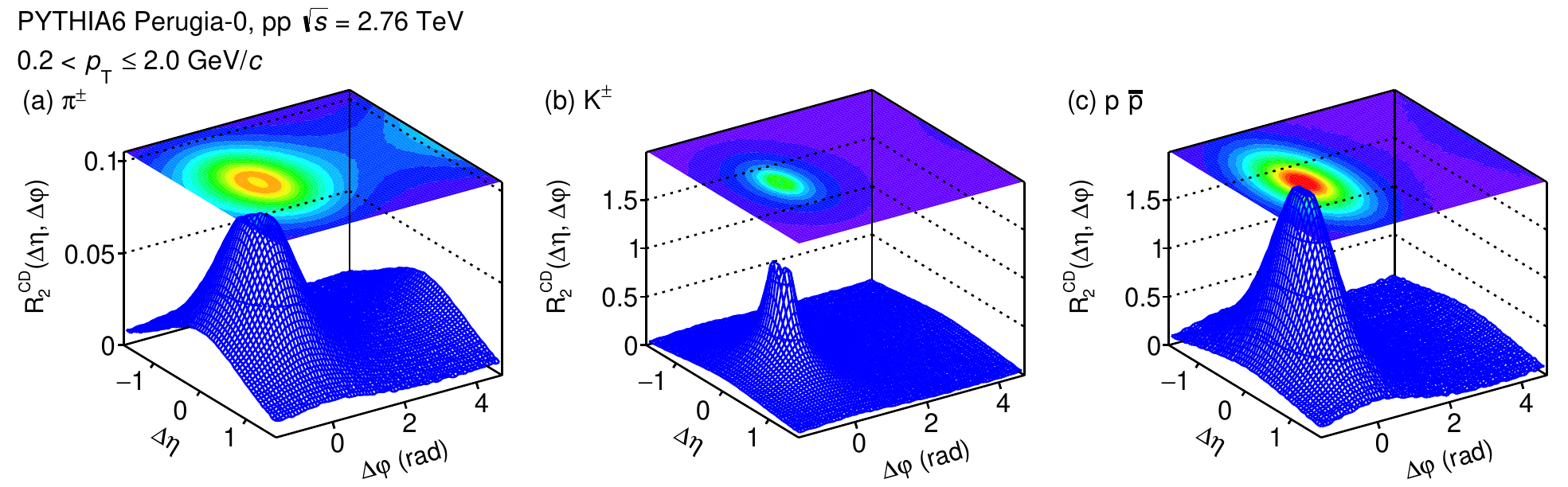}
 \caption{Correlation functions $\RtwoCD$ of $\pi^\pm$,
  $K^\pm$ and $p\bar{p}$, within $|\eta| < 1.0$
  and 0.2 < $\pt $ $\leq$ 2.0 $\gevc$, obtained with PYTHIA in \pp\
  collisions at $\s$ = 2.76 TeV.  }
 \label{fig:cR2CDidentified} 
\end{figure}
\begin{figure}[htb!]
 \includegraphics[scale=0.8]{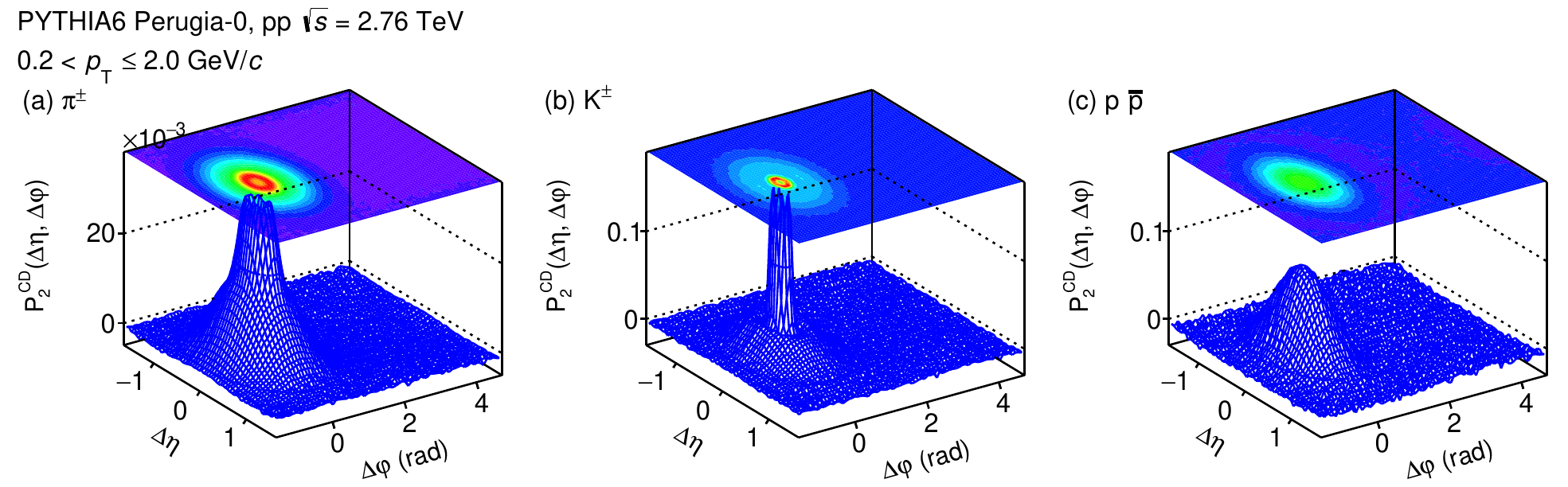}
 \caption{Correlation functions $\PtwoCD$ of $\pi^\pm$,
  $K^\pm$ and $p\bar{p}$, within $|\eta| < 1.0$
  and 0.2 < $\pt $ $\leq$ 2.0 $\gevc$, obtained with PYTHIA in \pp\
  collisions at $\s$ = 2.76 TeV.  }
 \label{fig:cP2CDidentified} 
\end{figure} 

\begin{figure}[ht!]
 \hspace*{-1.3cm}%
 \includegraphics[scale=0.3]{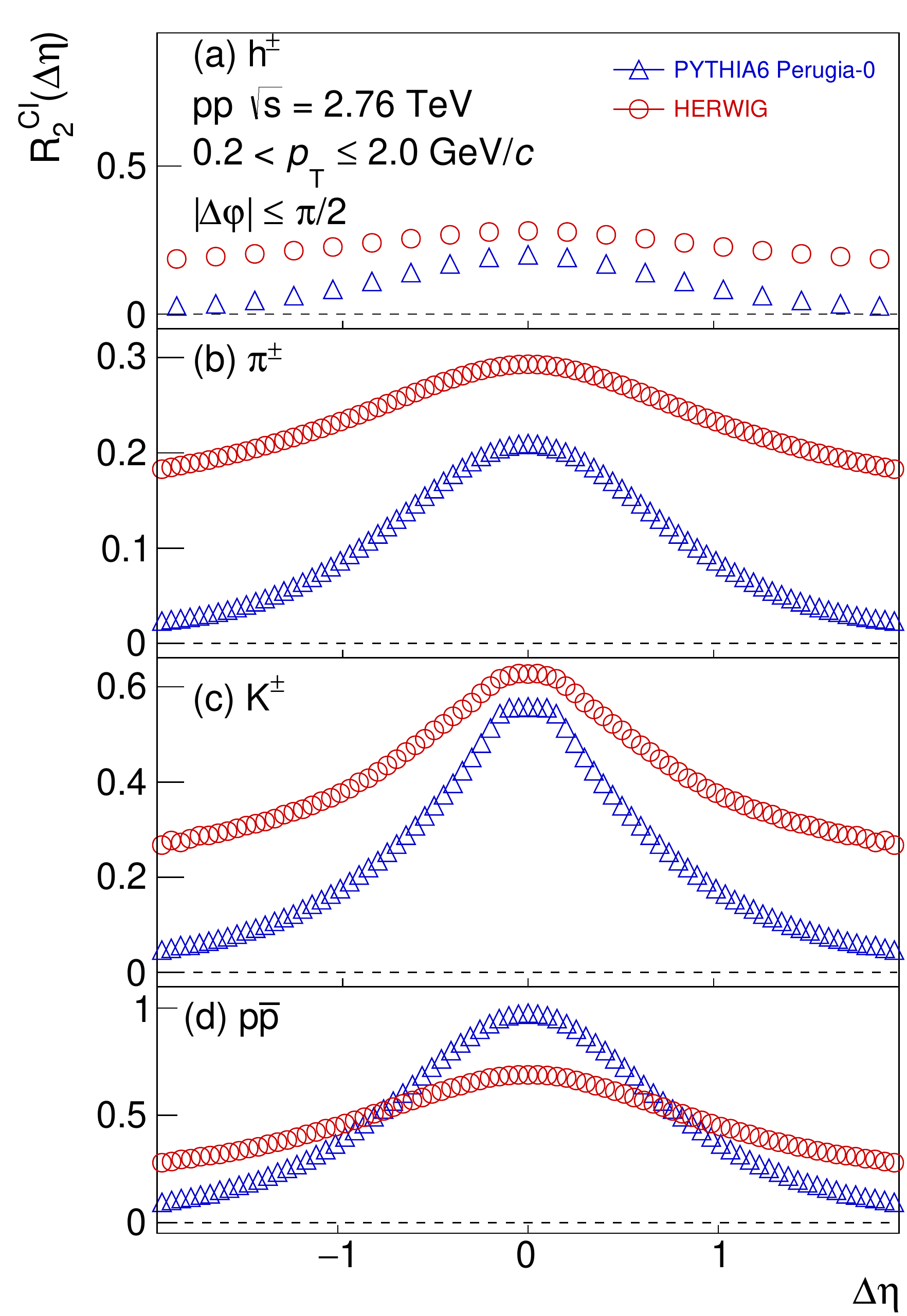}
 \includegraphics[scale=0.3]{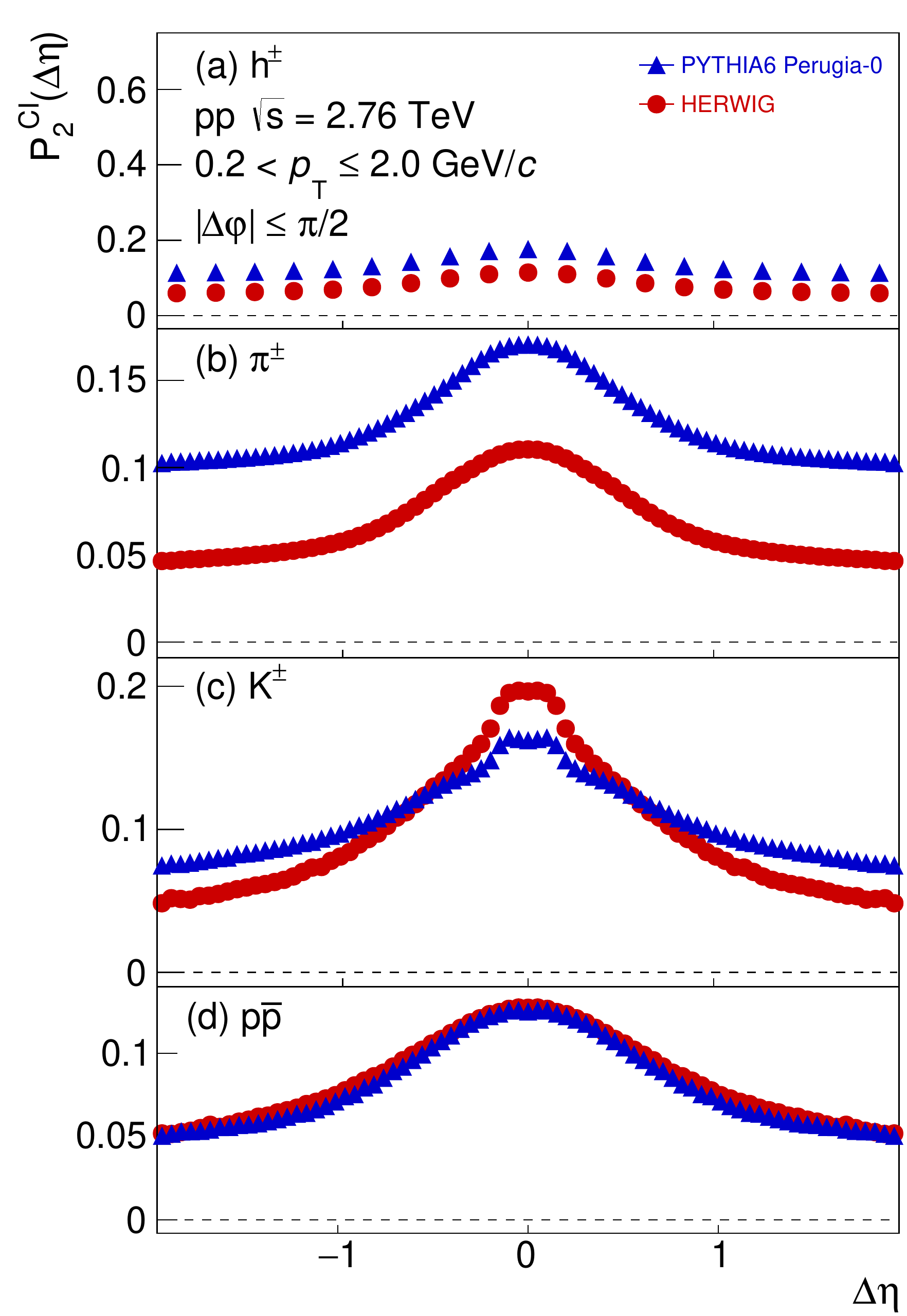}
 \caption{Projections onto $\Deta$ of $\RtwoCI$ (left column)
  and $\PtwoCI$ (right column) correlation functions of
  $h^\pm$, $\pi^\pm$, $K^\pm$ and $p\bar{p}$ calculated with PYTHIA (blue)
  and HERWIG (red) in \pp\ collisions at $\s$ = 2.76 TeV. The
  projections are calculated as averages of the two-dimensional
  correlations in the range $|\Dphi| \leq \pi/2$.}
 \label{fig:identifiedDetaCI}
\end{figure} 
\begin{figure}[ht!]
 \hspace*{-1.3cm}%
 \includegraphics[scale=0.3]{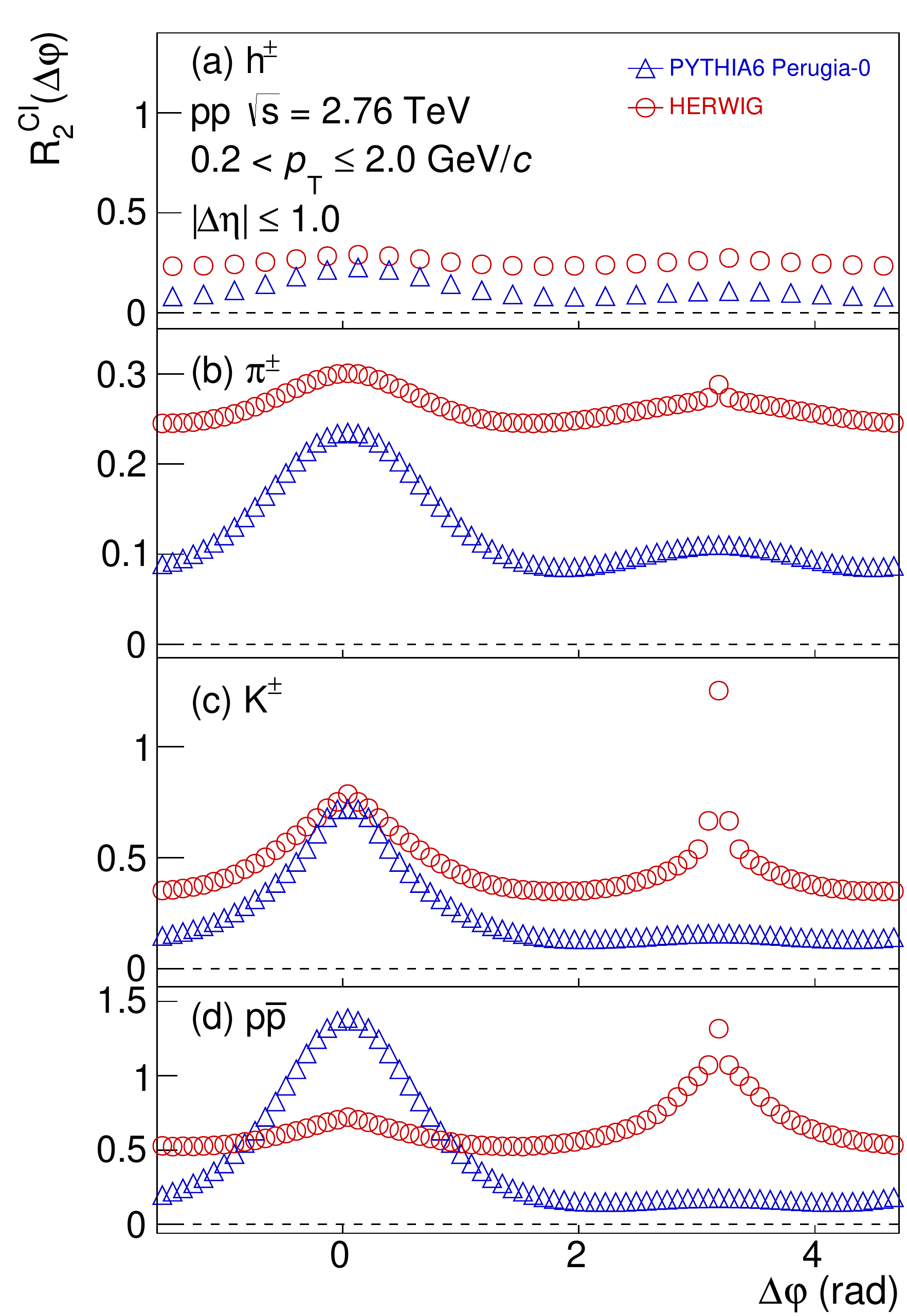}
 \includegraphics[scale=0.3]{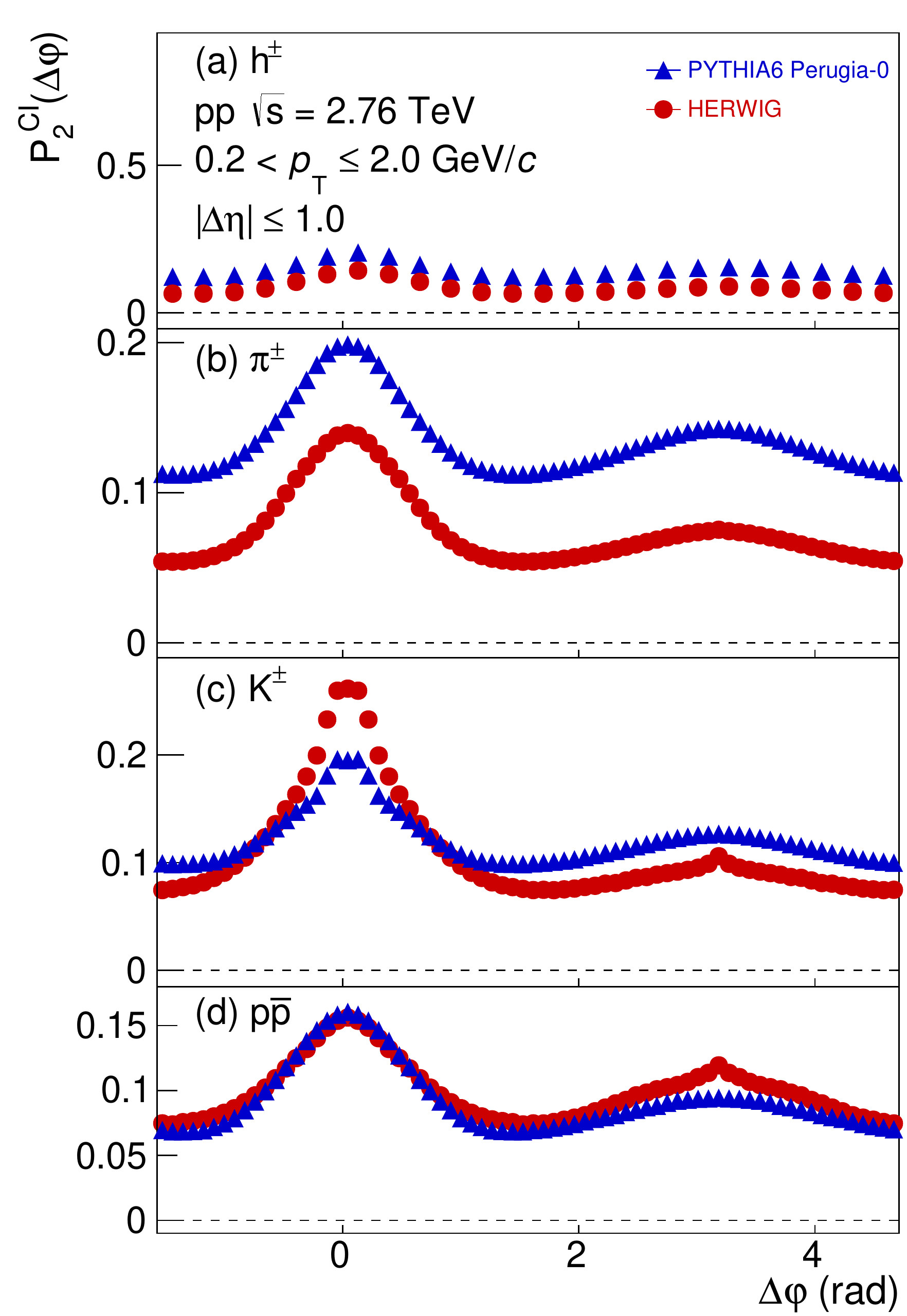}
 \caption{Projections onto $\Dphi$ of $\RtwoCI$ (left column)
  and $\PtwoCI$ (right column) correlation functions of
  $h^\pm$, $\pi^\pm$, $K^\pm$ and $p\bar{p}$ calculated with PYTHIA (blue)
  and HERWIG (red) in \pp\ collisions at $\s$ = 2.76 TeV.}
 \label{fig:identifiedDphiCI}
\end{figure} 

\begin{figure}[ht!]
 \hspace*{-1.3cm}%
 \includegraphics[scale=0.3]{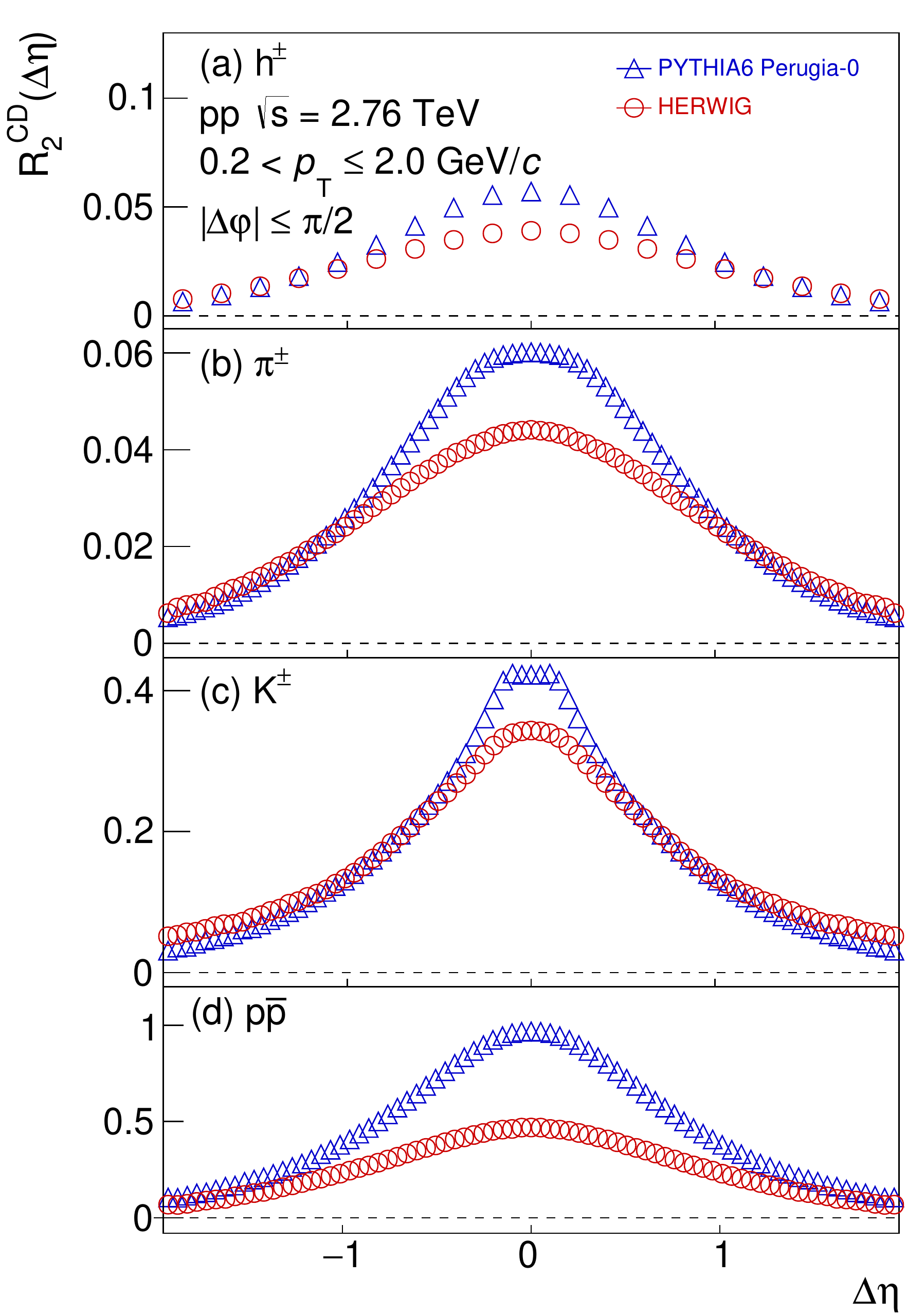}
 \includegraphics[scale=0.3]{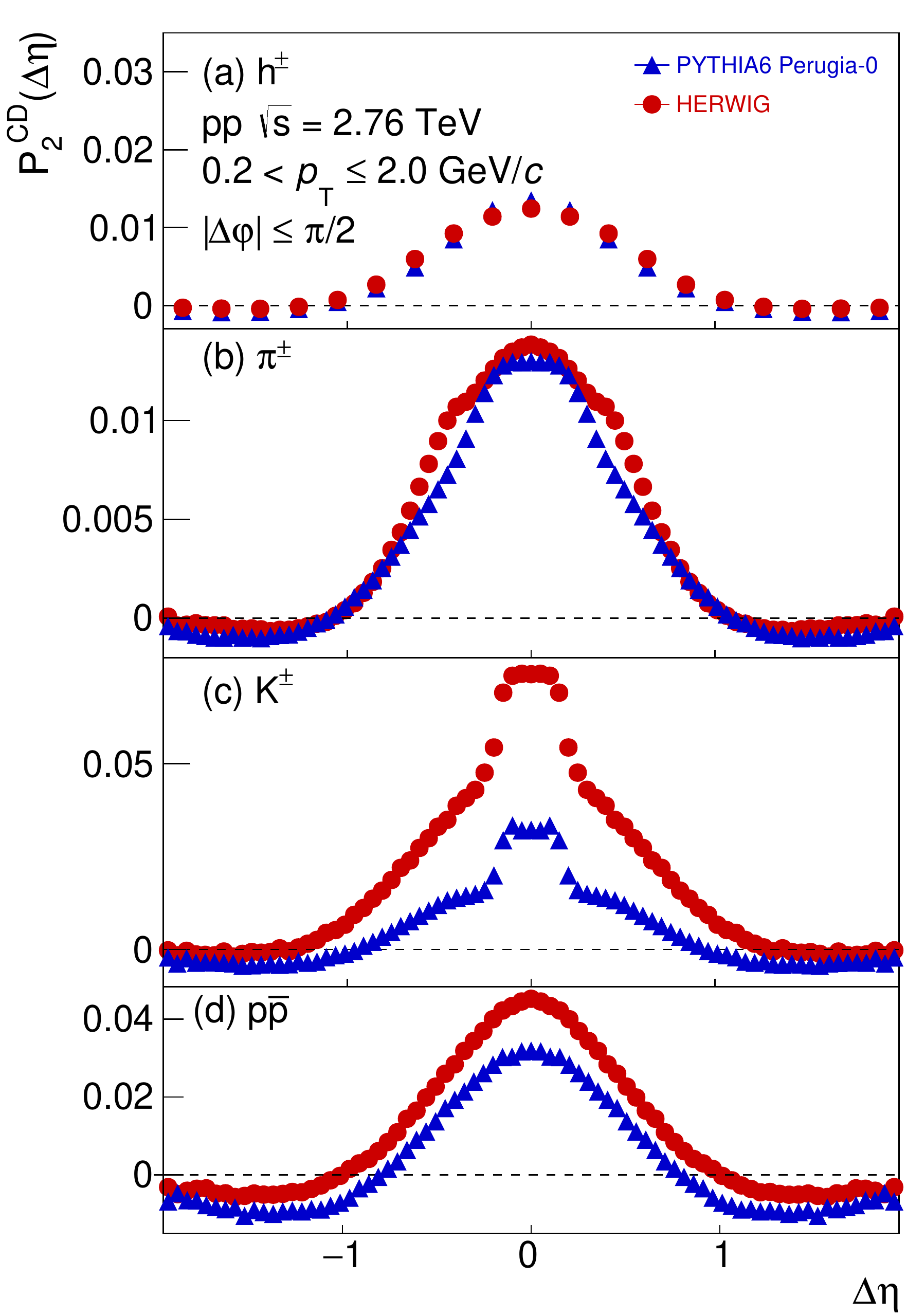}
 \caption{Projections onto $\Deta$ of $\RtwoCD$ (left column)
  and $\PtwoCD$ (right column) correlation functions of
  $h^\pm$, $\pi^\pm$, $K^\pm$ and $p\bar{p}$ calculated with PYTHIA (blue)
  and HERWIG (red) in \pp\ collisions at $\s$ = 2.76 TeV. The
  projections are calculated as averages of the two-dimensional
  correlations in the range $|\Dphi| \leq \pi/2$.}
 \label{fig:identifiedDetaCD}
\end{figure} 
\begin{figure}[ht!]
 \hspace*{-1.3cm}%
 \includegraphics[scale=0.3]{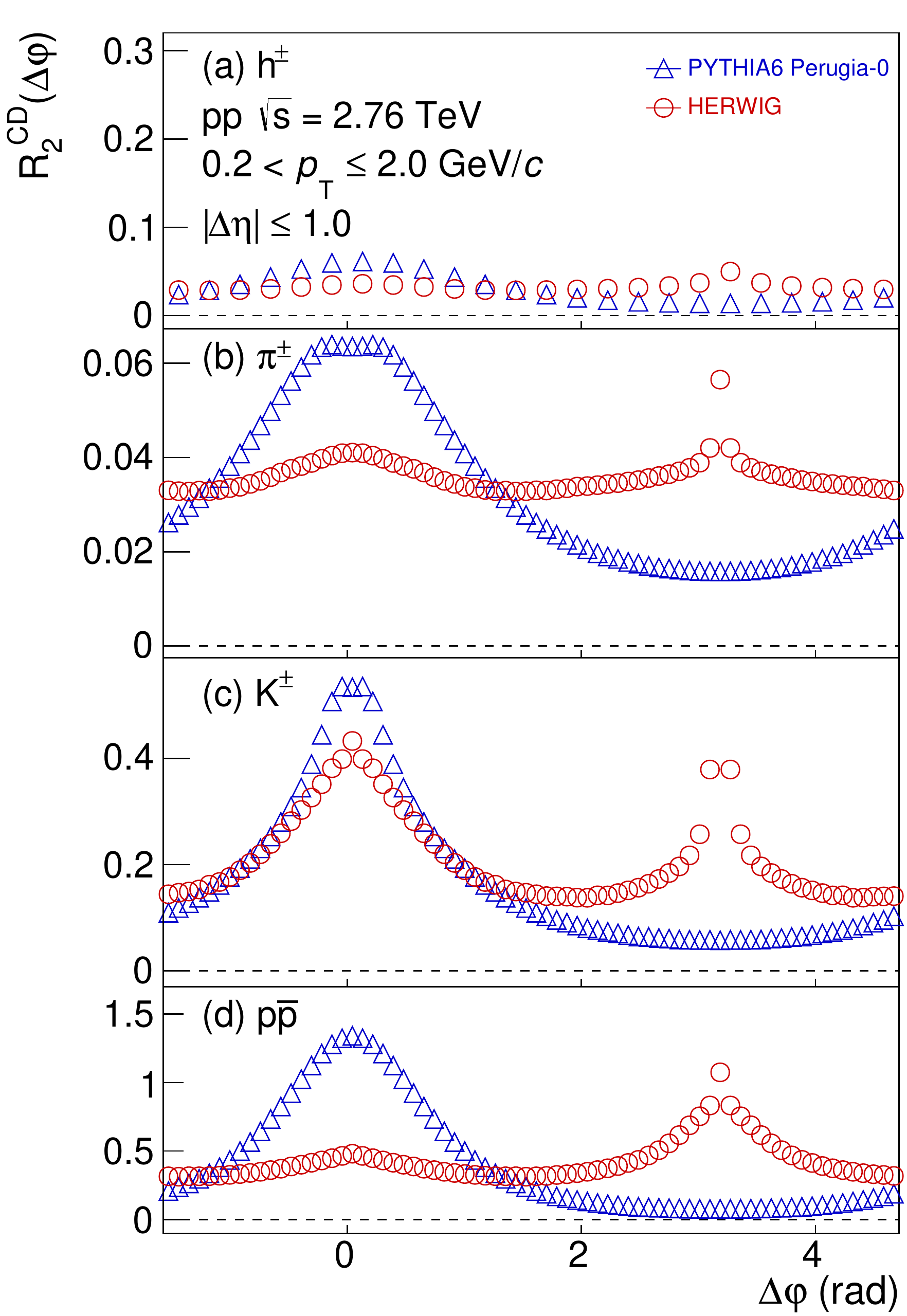}
 \includegraphics[scale=0.3]{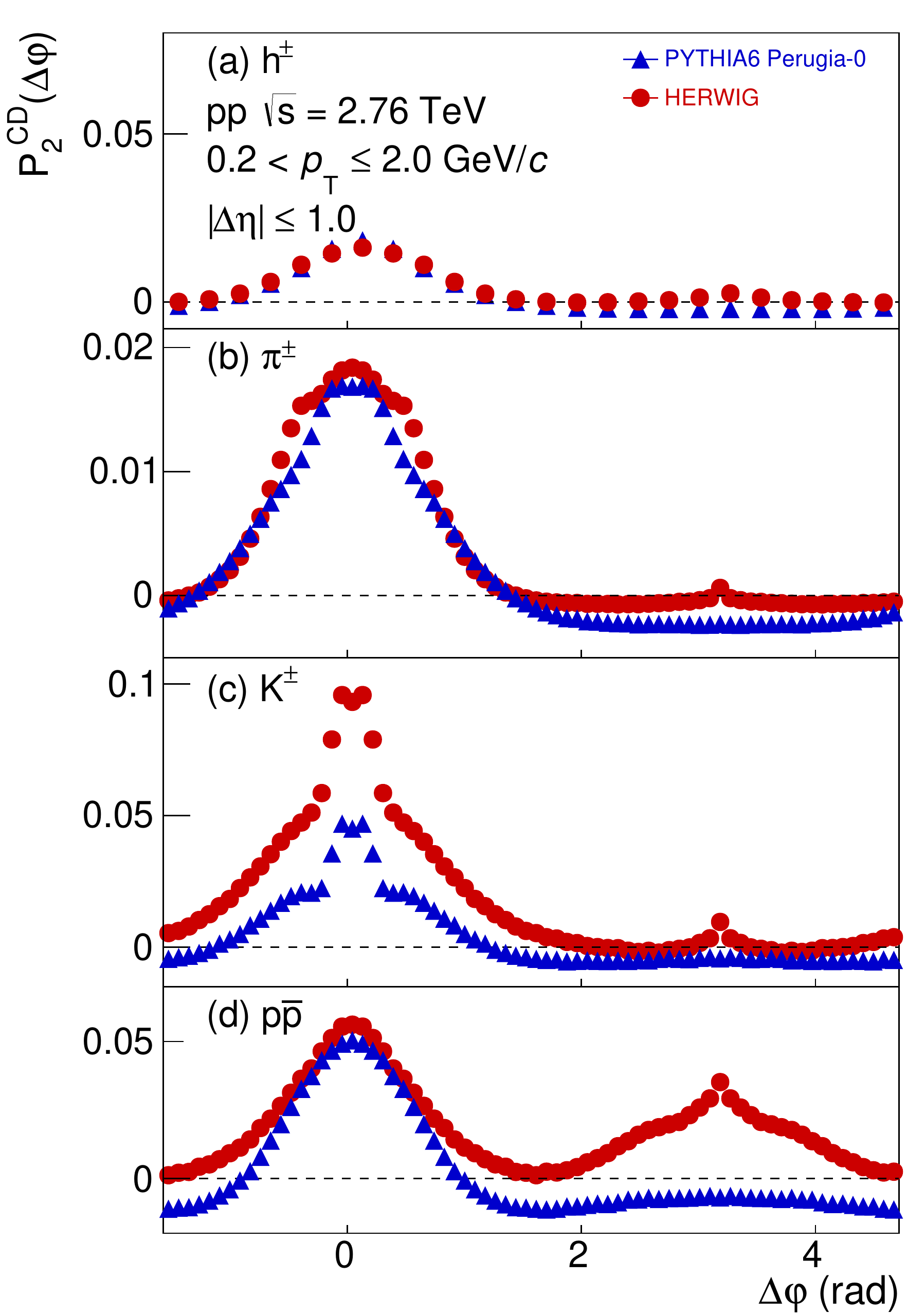}
 \caption{Projections onto $\Dphi$ of $\RtwoCD$ (left column)
  and $\PtwoCD$ (right column) correlation functions of
  $h^\pm$, $\pi^\pm$, $K^\pm$ and $p\bar{p}$ calculated with PYTHIA (blue)
  and HERWIG (red) in \pp\ collisions at $\s$ = 2.76 TeV.}
 \label{fig:identifiedDphiCD}
\end{figure}

Again in the case of
identified particles, one observes that the 
width of the near-side peak of the $\PtwoCI$ correlator is
significantly narrower than its $\RtwoCI$ counterpart. However,
the shape and width of these two correlators do not
exhibit a monotonic dependence on the mass of the particles. For
kaons, in particular, both $\RtwoCI$ and $\PtwoCI$ feature
a near-side peak that might be perhaps best described by a
superposition of a wide and a narrow Gaussian peak, which 
arises, in part, from a strong admixture of $\phi$-meson decays. A
similar situation arises for $\RtwoCD$ and $\PtwoCD$
shown in Figs.~\ref{fig:cR2CDidentified} and
\ref{fig:cP2CDidentified}, respectively. One finds, for all three
particle species, that the near-side peak of the $\PtwoCD$
correlators are markedly narrower than their $\RtwoCD$
counterparts. One also observes that the kaon near-side peaks are much
narrower than those of pions and protons.
It is also worth noticing that the pion
$\RtwoCD$ correlator shows a rather large away-side amplitude (relative
to its near-side peak amplitude) while kaons and protons feature much
smaller relative away-side amplitudes for this correlator.
By contrast, all three species have a
flat and nearly vanishing away-side amplitude in $\PtwoCD$
within PYTHIA simulations (Fig.~\ref{fig:cP2CDidentified}) for
particles within 0.2 < $\pt $ $\leq$ 2.0 \gevc. Qualitatively similar conclusions
are obtained from calculations of the $\Rtwo$ and $\Ptwo$ correlators with HERWIG
in this momentum range (2D plots not shown). Indeed, projections of the $\Rtwo$ and
$\Ptwo$ correlation functions obtained with PYTHIA and HERWIG, shown in
Figs.~\ref{fig:identifiedDetaCI}-\ref{fig:identifiedDphiCD}, illustrate
that while the predictions of the two models are qualitatively similar, 
they differ quantitatively for the three particle species considered. It is very difficult to study this in other $\pt$ regions for RMS width calculation because of large oscillating
behavior in $\PtwoCD$ $p\bar{p}$ in the $\pt$ range 2.0 - 5.0 $\gevc$. An actual
measurement of such correlation
functions (possible at the LHC with the ALICE detector) shall thus
provide significant constraints to  tune these models and achieve
a better understanding of particle production processes in elementary particle
collisions.

\section{Summary}\label{conclusion}

We presented a study of charge-independent and charge-dependent
two-particle differential- number correlation functions $\Rtwo$ and
transverse momentum correlation functions $\Ptwo$ in \pp\ collisions
at $\s$ = 2.76 TeV with the PYTHIA and HERWIG Monte Carlo
models. Calculations were presented for unidentified hadrons as well
as for $\pi^\pm$, $K^\pm$ and $p\bar{p}$ individual species in
selected ranges of transverse momentum. PYTHIA and HERWIG both
qualitatively reproduce the near-side peak and away-side ridge
correlation features reported by experiments. At low $\pt$, both
models produce narrower near-side peaks in $\Ptwo$ correlations than
in $\Rtwo$ as reported by the ALICE collaboration in \pPb\ and
\PbPb\ collisions~\cite{AliceDptDptLongPaper}. This suggests that the narrower shape of the $\Ptwo$ near-side peak
is largely determined by the $\pt$ dependent angular ordering of
hadrons produced in jets, as discussed in sec.~\ref{sec:definition}. We have
provided detailed calculations of the longitudinal and azimuthal widths
of the near-side peak as a reference to prospective experimental
studies of these correlation functions. Both PYTHIA and HERWIG 
predict widths that decrease with increasing $\pt$. Widths extracted
for $\Ptwo$ correlators are typically significantly narrower than those 
of the $\Rtwo$ counterparts. We also showed that the models
predict non-trivial dependence on the mass of identified particles
arising in part from resonance decays. 

 We additionally find that
the models produce large amplitude
ridge structures at $\Dphi=\pi$ in $\Ptwo$  
correlation functions while yielding relatively modest ridges in
$\Rtwo$. The amplitude of the ridge structure in $\PtwoCI$ is found
to increase with the particle $\pt$ range considered reaching rather
large amplitude for particles in the $5.0 < \pt \le 30.0~\gevc$ range.
An away-side ridge is also observed in $\PtwoCD$ correlation 
functions. The magnitude of this ridge  shall depend on jet-to-jet charge correlations. Measurements of $\PtwoCD$ correlation functions of high-$\pt$ particles in \pp\
collisions might then be sensitive to the charge of the partons
initiating the observed jets. Elucidation of this conjecture, however, requires further
studies, with both PYTHIA and HERWIG, of the correlation functions obtained when jet
production is restricted to gluon-gluon or quark-quark processes. 

\newenvironment{acknowledgement}{\relax}{\relax}
\begin{acknowledgement}
 \section*{Acknowledgements}
 
The authors thank Drs.~Joern Putschke and Sidharth Prasad for fruitful discussions
and their invaluable review of the manuscript.
This work was supported in part by the United States Department of
Energy, Office of Nuclear Physics (DOE NP), United States of America,
under grant No. DE-FOA-0001664, and the Department of Science and
Technology(DST), Government of India, under  grants No. SR/MF/PS-01/2014-IITB and
SR/MF/PS-02/2014-IITM, as well as the  University Grants Commission(UGC),
Government of India, under the Code No. BININ00401340.
\end{acknowledgement}

\bibliography{r2p2VsModels}
\end{document}